\documentclass[12pt]{iopart}

\usepackage{amsopn}
\usepackage{iopams}
\usepackage{setstack}
\usepackage{graphicx}
\usepackage[T1]{fontenc}
\usepackage[english]{babel}
\usepackage[pdfpagelabels,plainpages=false]{hyperref}   
\usepackage{amssymb}
\usepackage{fancyhdr}
\usepackage{units}
\usepackage{dsfont}
\usepackage{braket}
\usepackage{color}      
\newcommand{\unity}{\ensuremath{\mathds{1}}}

\newcommand{\Fig}[1]{figure~\ref{#1}}
\newcommand{\Sec}[1]{section~\ref{#1}}
\renewcommand{\S}{\hat{S}}

\newcommand{\skippart}[1]{}
\newcommand{\vmax}{v_{\rm max}} 

\DeclareMathOperator{\Ai}{Ai}
\DeclareMathOperator{\dn}{dn}
\DeclareMathOperator{\sn}{sn}
\DeclareMathOperator{\cn}{cn}

\def\figpath{.} 
\def\appendixfigpath{.} 

\begin{document}

\title[Time Evolution within a Comoving Window]{Time Evolution within a Comoving Window: Scaling of signal fronts and magnetization plateaus after a local quench in quantum spin chains}
\author{V~Zauner$^{1,2,3}$, M~Ganahl$^{1,4}$, H~G~Evertz$^{1}$, T~Nishino$^{3}$}

  \address{$^{1}$Institute of Theoretical and Computational Physics, Graz University of Technology, 8010 Graz, Austria}
  \address{$^{2}$Vienna Center for Quantum Technology, University of Vienna, 1090 Vienna, Austria}
  \address{$^{3}$Department of Physics, Graduate School of Science, Kobe University, Kobe 657-8501, Japan}
  \address{$^{4}$Perimeter Institute for Theoretical Physics, Waterloo, ON N2L 2Y5, Canada}
  
\ead{valentin.zauner@univie.ac.at}

\begin{abstract}
We present a modification of Matrix Product State time evolution to simulate the propagation of signal fronts on
infinite one-dimensional systems. We restrict the calculation to a window moving along with a signal,
which by the Lieb-Robinson bound is contained within a light cone.
Signal fronts can be studied unperturbed 
and with high precision
for much longer times than on finite systems. 
Entanglement inside the window is naturally small, greatly lowering computational effort. 
We investigate the time evolution of the transverse field Ising (TFI) model and of the $S=1/2$ XXZ antiferromagnet 
 in their symmetry broken phases
after several different local quantum quenches.

In both models, we observe distinct magnetization plateaus at the signal front for very large times,
resembling those previously observed for the particle density of tight binding (TB) fermions. 
We show that the normalized difference to the magnetization of the ground state exhibits 
similar scaling behaviour as the density of TB fermions.
In the XXZ model there is an additional internal structure of the signal front due to pairing,
and wider plateaus 
with tight binding scaling exponents for the normalized excess magnetization.
We also observe parameter dependent interaction effects between individual plateaus, resulting in a slight spatial compression of the plateau widths.

In the TFI model, we additionally find that for an initial Jordan-Wigner domain wall state, 
the complete time evolution of the normalized excess longitudinal magnetization 
agrees exactly with the particle density of TB fermions. 
\end{abstract}

\noindent{Keywords: Strongly Correlated Systems, Tensor Network States, Quantum Quenches, Unitary Time Evolution, Particle Propagation, Integrable Systems

\submitto{\NJP}

\maketitle

\section{Introduction}
Signal propagation in one-dimensional (1D) strongly interacting quantum lattice systems has been 
of longstanding general interest in both condensed matter and quantum-computational physics,
where it provides a basis for coherent information transfer via quantum wires. 
A signal can be created, e.g., as a local excitation from a stationary state, or as a domain wall or a topological
excitation~\cite{Cheneau12,d_wall}.  Often hard to pursue by analytical methods, many studies have become feasible in 1D 
due to Matrix Product State (MPS)~\cite{FCS,MPS1,MPS2} based numerical methods~\cite{idmrg1,tebd,tdmrg}.
Thus the non-equilibrium time evolution of such signals
after global
\cite{Calabrese07,Calabrese11,IgloiRieger00, Igloi11,Rieger11,Eisler09,Rossini09,Rossini10,Foini11,Schuricht12,Barmettler0910,Fagotti13,Fagotti14,Pollmann13,Brockmann14}
and local 
\cite{Karevski02,Platini05,Eisler08,Divakaran11,Gobert05,Pereira08,Langer0911,Lancaster10,Mossel10,Foster11,Jesenko11,Ganahl12,Alba14,Karrasch14,Sabetta13,Halimeh14,Abraham70,Antal99,Hunyadi04,Antal08}
quantum quenches has been the subject of intense theoretical interest in recent years.
%
In particular for tight binding (TB) fermions initially in a domain wall (DW) state, intriguing plateaus in the fermion density have been found to develop at large times with well defined scaling behaviour
\cite{Antal99,Hunyadi04,Antal08}
and have only been fully understood recently \cite{Eisler13}.
%

If an initial state for such a study is prepared within a \textit{finite}  
system, boundary effects such as Friedel oscillations interfere with a passing signal. 
System boundaries also limit the time span for signal tracing before non-trivial reflections occur at the boundaries.
The maximum time is even more severely restricted by entanglement 
which develops across the system and which requires a computational effort that can drastically increase with
time~\cite{blowS,Prosen07}.
This has greatly hampered the analysis of large time asymptotic behaviour
\cite{Barmettler0910,Gobert05,Jesenko11}.
Boundary effects do not appear in \textit{infinite} systems, 
for which the ground state and its time evolution can be efficiently calculated with MPS 
methods \cite{idmrg1,iTEBD,idmrg2,TDVP}.
%
However, these methods require complete translation invariance and can therefore not be applied to study signal
propagation.


In this paper we present a simple method to simulate the propagation of local signals on an infinite chain
using MPS time evolution, without any finite size effects distorting the signal front.
For related approaches to boundary effects, see \cite{phien-infinite,phien-moving,Osborne,preprints}.
We study the time evolution of the Transverse Field Ising (TFI) model 
and of the spin-$1/2$ XXZ chain
after \textit{local} quantum quenches up to large times, which were not accessible before using conventional MPS techniques.
In both models we observe distinct magnetization plateaus 
developing over time close to the signal front similar to the case of TB fermions \cite{Antal99,Hunyadi04,Eisler13}, and which also exhibit similar asymptotic scaling.
Surprisingly we find an exact agreement at all times and positions between the magnetization in the TFI model and the density of TB fermions for a particular type of signal.
For the XXZ chain we observe interaction effects between individual plateaus, which can be tuned via the model parameters.

For our method we consider a spin chain of infinite size with nearest neighbour interactions, 
initially prepared in a state -- such as the ground state -- which is translation invariant
for sites $n>n_0$ to the right of some site $n_0$.
At time zero, the system is excited by a quantum quench 
like one or more spin flips at sites $n\le n_0$ 
or a modification of the Hamiltonian at $n\le n_0$. 
For local interactions 
it is known from the Lieb-Robinson bound \cite{lieb-rob,barthel12}
that wave fronts generated by such quenches can at most propagate with a
characteristic maximum velocity $\vmax$, 
i.e. within a  "light cone" even in a non-relativistic system
as recently also seen experimentally \cite{Cheneau12}.
Any correlations beyond the light cone
are exponentially suppressed. 
In the following  we will consider right moving signals for the sake of concreteness.

\section{Method}
Our approach is to introduce a division of the system into three parts, namely a \textit{comoving window} (CMW) -- which moves towards the
right with the wave front -- and two half-infinite parts, a uniform one in front (i.e. to the right) of the window, and an
arbitrary one to the rear. 
The window is chosen wide enough to contain the complete signal front, 
including the exponentially damped part to the right of the main front, to high precision.
The signal therefore does not affect the uniform system to the right of the window.
Likewise, when the window moves with $\vmax$,
modifications in the rear 
part do not affect the CMW and need not be calculated. The method is therefore fit for studying fronts of propagating signals, 
in particular those generated by local quenches.
Since bipartite entanglement \cite{Eisler09,Eisler08,amico04,Amico08}
spreads at most with $\vmax$, 
the bipartite entanglement entropy is significantly lower around the wave front than in the bulk,
allowing for reduced computational effort when using the CMW.

We mark the left and right boundary of the CMW with indices $\ell$ and $r$ respectively and divide the system into
left part $j \le \ell$,  CMW $\ell + 1 \le j \le r$, and right part $j\ge r+1$.  
The Hamiltonian  
$\hat H =  \sum_j \hat h_{j,j+1}$
subdivides correspondingly into
\begin{equation}
{\hat H} = {\hat H}_{\rm L}^{~} + {\hat h}_{\ell, \ell + 1}^{~} + {\hat H}_{\rm M}^{~} 
+ {\hat h}_{r, r+1}^{~} + {\hat H}_{\rm R}^{~} \, .
\end{equation}
Low energy states of the overall system are well approximated by Matrix Product States (MPS) \cite{MPS1,MPS2} 
and we write the wave function as an MPS in the so-called mixed canonical form as
\begin{equation}
\psi(\{s_{j}\})=
\ldots L^{s_{\ell-1}} L^{s_{\ell}}
A^{s_{\ell+1}}\ldots A^{s_k}\lambda^{k}
B^{s_{k+1}}\ldots B^{s_r} 
R^{s_{r+1}}_{~} R^{s_{r+2}}_{~} \ldots ,
\label{eq:CMW_MPS}
\end{equation}
%
where $s_{j}$ labels the spins, $L^{s_{j \le \ell}}$ are left-orthogonal matrices ($\sum_{s_\ell}{L^{s_j}}^{\dagger}L^{s_j}=\unity$)
defined on the left part, $R^{s_{r+1 \le j}}$ are right-orthogonal matrices
($\sum_{s_j}R^{s_j}{R^{s_j}}^{\dagger}=\unity$) defined on the right part,
$A^{s_{\ell +1 \le j \le r}}$ and $B^{s_{\ell +1 \le j \le r}}$ are left- and right-orthogonal matrices respectively defined inside
the CMW, and $\lambda^{\ell\le k\le r}$ are diagonal matrices containing the Schmidt values of a
bipartition at bond $(k,k+1)$. 
For a finite system, the left and right ends of \eref{eq:CMW_MPS} are terminated by contractions
with boundary vectors; we however consider the infinite size limit.

\begin{figure}[tb]
 \centering
 \includegraphics[width=\linewidth,keepaspectratio=true]{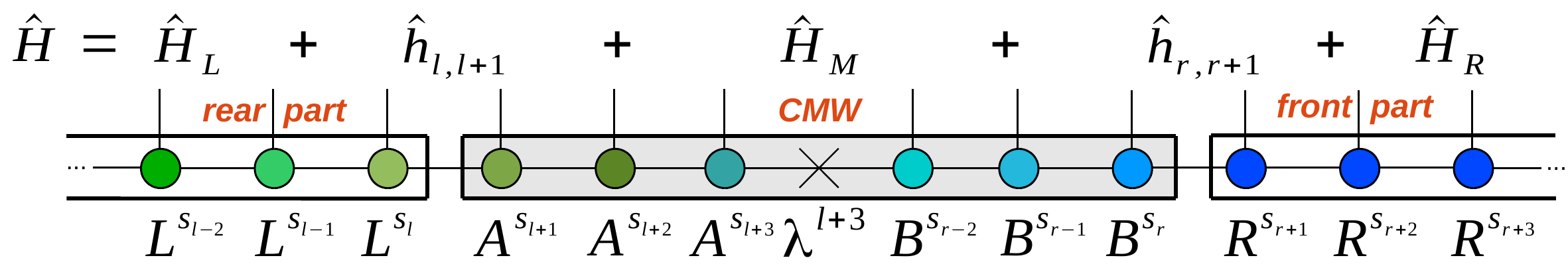}
 \caption{Graphical representation of the MPS describing the overall system state, which is divided into comoving window (CMW), left and
right part.}
 \label{fig:CMW_MPS}
\end{figure}

The matrices $R^{s_j}$ describe the uniform half-infinite system in the front and are therefore constrained to be
translation invariant. We use a 2-site unit cell, i.e. $R^{s_{j+2}}=R^{s_{j}}$. The matrices $A^{s_{j}}$
and $B^{s_{j}}$ describe the CMW and are site dependent. For the matrices $L^{s_{j}}$, which describe the left
part, we impose no uniformity restrictions. 
They represent initial conditions for the left boundary of the CMW and remain unchanged throughout the
simulation. Additional matrices are added to this collection of $L^{s_{j}}$ whenever the CMW is moved.

Let us consider one step of unitary time evolution for the entire system. Inside the CMW, between sites 
$\ell+1$ and $r$, 
we employ time dependent Density Matrix Renormalization Group (tDMRG \cite{tdmrg}), using a second order even-odd Suzuki-Trotter decomposition~\cite{TrotterSuzuki} with 
local operators $\hat{u}_{j,j+1}( \tau ) = \rme^{-\rmi \tau \hat{h}_{j,j+1}}$ and small time steps $\tau$. 


In order to connect time evolution inside and outside the CMW we introduce two different approaches, which we now
sketch for the case of the right (front) and the left (rear) boundary respectively. Details can be found in \ref{sec:timeevo}.

In \textit{Method I (Uniform Update)}, applied to the right boundary, the matrices $R^{s_{j}}$ of the right
part are first updated by infinite system Time Evolving Block Decimation (iTEBD \cite{iTEBD}). 
We then evolve the junction bond $(r,r+1)$ by applying $\hat{u}_{r,r+1}$ and we exploit right-orthogonality of $R^{s_{r+1}}$ to update $B^{s_{r}}$ and
to ensure gauge consistency of MPS matrices around the junction bond.

For \textit{Method II (Renormalized Update)}, applied to the left boundary, we adapt the algorithm of Cazalilla and Marston 
\cite{non_adapt} (Method II is similar to the algorithm introduced in \cite{phien-infinite,phien-moving}, see \cite{preprints})
and construct a renormalized representation for 
$H_{\rm L} + h_{\ell,\ell+1}$ to approximate the evolution of the left part and the left junction bond $(\ell,\ell+1)$, 
such that  all changes in the left part are compressed into the boundary matrix
$A^{s_{\ell+1}}$, and the matrices $L^{s_{j\leq\ell}}$ remain unchanged.

The Uniform Update
has some immediate advantages. It is easier to implement 
and it is also applicable in case of 
a time dependent $H_{R}$. It does however require translation invariance of the right part.
%
The Renormalized Update
does not preserve the structure of the Suzuki-Trotter decomposition at the boundary
and therefore continually introduces small perturbations there.
In 
\ref{sec:precision} 
we 
compare both methods to analytical results and to a reference
  simulation on a very large stationary lattice and 
show that both methods work well.
As errors in our new Uniform Update, when applied to the right boundary, 
are only of order $\Or(10^{-8})$ 
and thus smaller  
by several orders of magnitude than for the Renormalized Update, we use the Uniform Update for the right boundary.

For the left boundary, the simplest approach is to disconnect the left part by setting $\hat h_{\ell,\ell+1}=0$, which already works quite well (see \ref{sec:precision})
when the window moves with $\vmax$, as then any perturbations are confined to the neighbourhood of the rear boundary.
Since perturbations there are however smallest with 
the Renormalized Update, we use this method
for the left boundary in the present paper, 
For further details on the boundary updates and how to move the CMW along with a propagating signal see \ref{sec:timeevo}.

\section{Results}
\subsection{Transverse Field Ising (TFI) model}

 \begin{figure}[tb]
 \centerline{\includegraphics[width=\textwidth,clip]{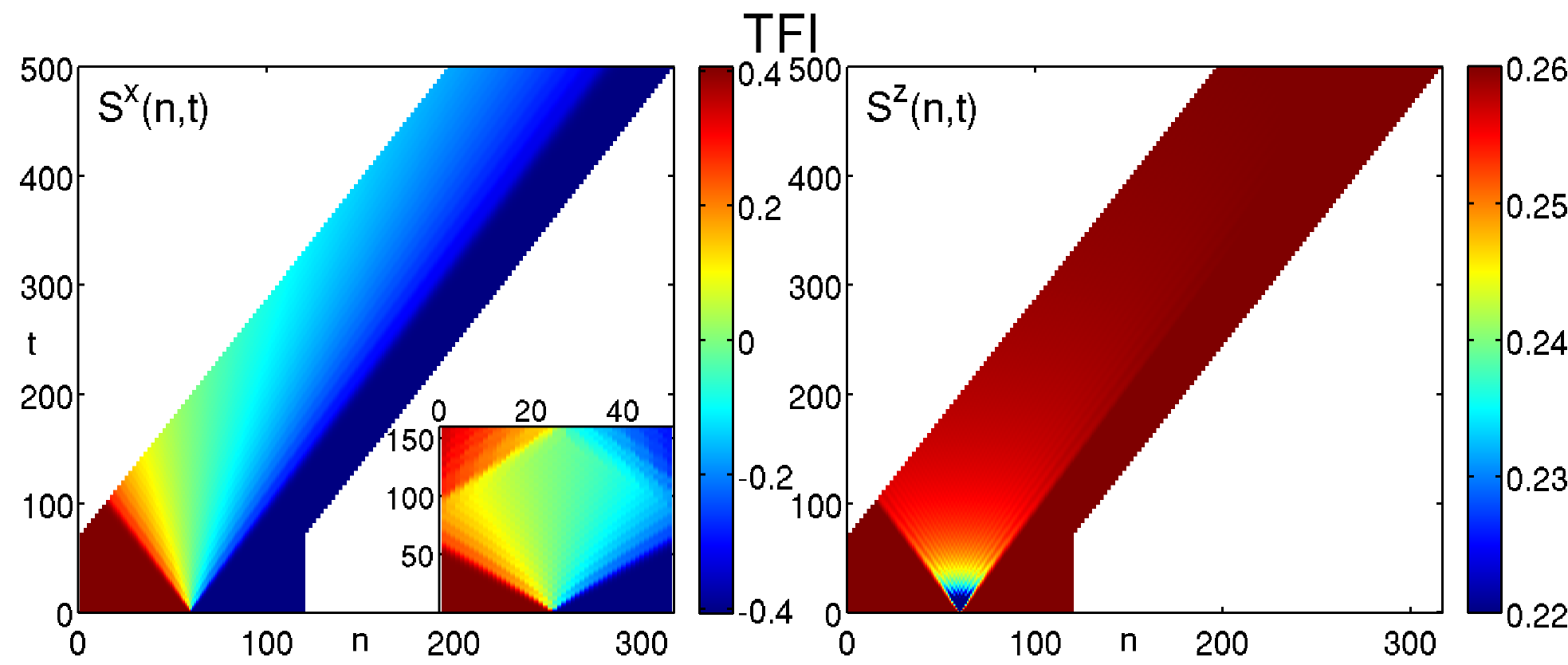}}
 \caption{
 Time evolution of magnetizations $S^x(n,t)$ and $S^z(n,t)$ in the TFI model at $h = 0.45$
 after a JW excitation.
 We show times only up to $t=500$ in order to keep structures resolvable to the eye, while simulations were performed up to $t=1000$.
 \textit{Inset:} Time evolution of $S^x(n,t)$ without window movement, showing eventual reflections.
 }
 \label{fig:TFI_TimePlot}
 \end{figure}

\begin{figure}[tb]
 \centering
 \includegraphics[width=0.9\linewidth,keepaspectratio=true]{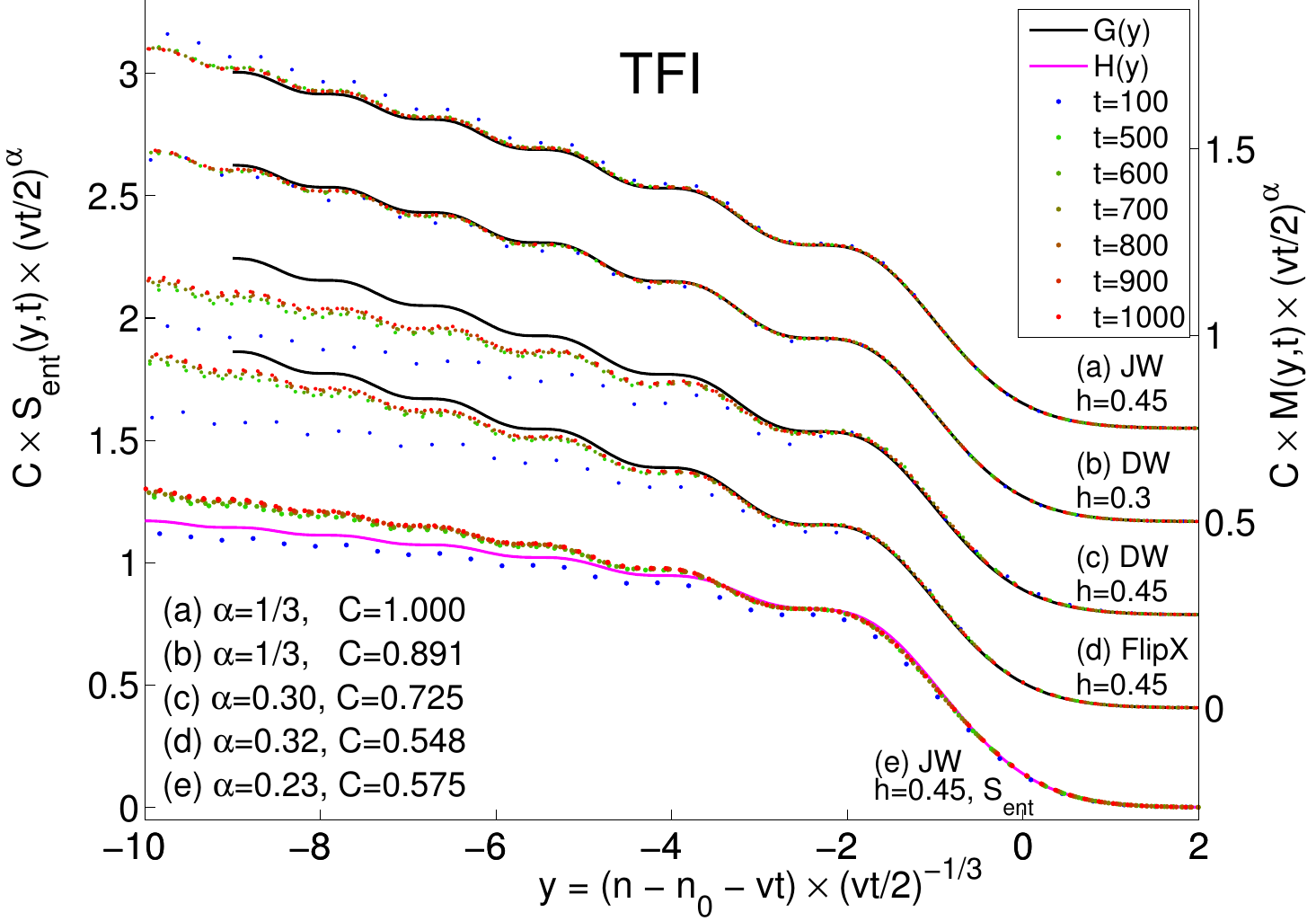}
 \caption{\textit{Scaled} normalized excess magnetization $M(y,t)$
(right axis) and bipartite entanglement entropy $S_{\rm ent}(y,t)$
(bottom only, left axis) vs. scaled position $y$ at the signal front for the TFI model at different $h$ for various signal types.
$G(y)=[\Ai'(y)]^2 - y\Ai(y)^2$ and $H(y)$ are the density and entropy scaling functions for TB fermions \cite{Eisler13}.
The lines are successively offset by $0.25$ in vertical direction.
}
\label{fig:TFI_combined}
\end{figure}
The spin-1/2 TFI model
\cite{Calabrese07,Calabrese11,IgloiRieger00, Igloi11,Rieger11,Eisler09,Rossini09,Rossini10,Foini11,Schuricht12,Karevski02,Platini05,Eisler08,Divakaran11}
on an infinite chain defined by
\begin{equation}
\hat{H}=-\sum_n \hat{S}^x_n \hat{S}^x_{n+1} - h\sum_n \hat{S}^z_n
\label{eq:TIM_ham}
\end{equation}
can be solved exactly \cite{lsm,Pfeuty70} (see also \ref{sec:TFI_analytics}),
and the time evolution of local observables
can in principle be calculated \cite{IgloiRieger00,Calabrese11}.
For the \textit{longitudinal} magnetization $S^x(n,t)$ (order parameter), 
analytical calculations are however difficult and
some results have become available in the literature only recently \cite{Calabrese11,Igloi11},
but to our knowledge not for local quenches on infinite systems.
In the ferromagnetic phase $h<h_c=0.5$ 
the ground state is twofold degenerate and there is long range order in $S^x$.

We prepare the system in the maximally symmetry broken ground state $\ket{\Downarrow}$ (\ref{sec:init})
with $S^{x}_{\rm GS}:=\braket{\S^x_n} <0$ using iDMRG \cite{idmrg1,idmrg2} and study the time evolution of several 
initial states excited from $\ket{\Downarrow}$.
In \Fig{fig:TFI_TimePlot} we show results for a Jordan-Wigner (JW) excitation
\begin{equation}
 (c^\dagger_{n_0} + c_{n_0} ) \ket{\Downarrow}
 = \prod_{n<{n_0}} (-2\hat{S}^z_n )( 2\hat{S}^{x}_{n_{0}})\ket{\Downarrow}
 \label{eq:JW_excitation_TFI}
\end{equation} 
on site $n_0$  inside the window, 
where $c^\dagger,c$ are JW fermion 
operators (see \cite{jw} and \ref{sec:TFI_analytics}). 
This corresponds to a spin flip in $z$-direction at site $n_0$ and 
a domain wall in $x$-direction between sites $n_0-1$ and $n_{0}$. 
%
Window movement is triggered by bipartite entanglement entropy, resulting in window velocities consistent with exact maximum velocities (\ref{sec:TFI_analytics}). 
We use a
second order Suzuki-Trotter decomposition with a step size of $\tau=0.002$ and maximum matrix dimension $m_{\rm max}=120$ during time evolution.
%
The time evolution inside the CMW (\Fig{fig:TFI_TimePlot})
shows that boundary effects are indeed removed at both ends of the CMW. 
In \ref{sec:precision} we show that 
results inside the CMW are unperturbed to very high accuracy (about $10^{-8}$) at all times.

When the window is not moved (\Fig{fig:TFI_TimePlot}, inset),
the signal is absorbed by both boundaries temporarily, but reflections emerge eventually
with both methods.
This remains true also for additional models studied in \ref{sec:reflections}, in all cases.
%
We also investigate a pure domain wall (DW) excitation 
$ \prod_{n<{n_0}} (2\hat{S}^{z}_{n})\ket{\Downarrow}$ between sites $n_0-1$ and $n_{0}$
and a spin flip  in $x$-direction (FlipX) $(2\hat{S}^z_{n_{0}} )\ket{\Downarrow}$ at site $n_{0}$.

\subsubsection{Step structure.}
Despite different global shapes (see \ref{sec:unscaled_results}) for the different excitations, we find
that a step structure always develops in $S^{x}(n,t)$ at the signal front at large times (\Fig{fig:TFI_combined}), 
similar to the time evolution from an initial DW state for TB fermions \cite{Antal99,Hunyadi04,Eisler13}. 
The step structure takes much longer to develop for FlipX and DW excitations than for the JW case.
%
The transverse magnetization $S^z(n,t)$ does not show such a step structure.

The step structure is expected to be related to the ballistic nature of propagation at the signal front \cite{Alba14,Karrasch14,Eisler13},
like for TB fermions, where the steps are now fully understood as individual propagating particles \cite{Eisler13}.
For the TFI model, in different quench scenarios where two initially separate chains are joined,
beginnings of steps have previously been visible in results of \cite{Divakaran11}, but have not been investigated further.
We are not aware of other occurrences for the symmetry broken phase.
In the paramagnetic phase at large $2h=10$,
TB-like scaling has been observed in \cite{Platini05}
for the \textit{transverse} magnetization $S^{z}(n,t)$ after joining two initially separate chains at different temperatures. No steps occurred for the longitudinal magnetization.
Due to their quantum origin these steps appear not to be accessible \cite{Eisler13,RiegerComm} by semi-classical approaches such as in \cite{Rieger11}.

We find that the proper quantity to analyze our results is the normalized excess longitudinal magnetization
 \begin{equation}
M(n,t) \equiv [S^{x}(n,t)-S^{x}_{\rm GS}]/{|2S^{x}_{\rm GS}|}  ~.
\label{eq:excess_mag}
  \end{equation} 
Figure \ref{fig:TFI_combined} shows that at large times this quantity
indeed obeys the same scaling behaviour as the particle density of TB fermions \cite{Eisler13}
at the signal front.
For the DW and FlipX cases, there is  an additional proportionality factor $C\neq1$.
The asymptotic scaling function $G(y)$ for TB fermions \cite{Eisler13} is approached from different directions for different excitations.
For DW and FlipX excitations, the exponent $\alpha$ with best data collapse depends on $h$, 
whereas for the JW case it is independent of $h$. 

\subsubsection{Exact identity.}
In fact, for the JW excitation we find a surprising much closer identity with TB fermions:
The complete time evolution of the normalized excess longitudinal magnetization obeys
\begin{equation}
M(n,t)=N_{\rm TB}(n,vt) 
\end{equation}
 where $v=h$ is the TFI signal 
velocity (\ref{sec:TFI_velocity}) and 
$N_{\rm TB}(n,vt)$ is
the particle density of TB fermions at time $vt$ 
after a DW excitation
(steplike initial density as in \cite{Eisler13}).
%
We find this identity to hold 
  up to the numerical precision of our data
for all sites $n$ and times $t$ for $h<h_{\rm c}$, i.e. in the ferromagnetic phase,
but for the longitudinal magnetization only.

The steps in $N_{\rm TB}(n,t)$ have been shown to correspond to individual propagating particles \cite{Hunyadi04,Eisler13}
and we note that in the case of the TFI model a similar interpretation in terms of individual quasi-particles can only be given 
to the scaled excess longitudinal magnetization $M(n,t)$ after a JW excitation in the ferromagnetic phase. 
Due to the twofold degeneracy of the ground state in this phase the application of a \textit{local} perturbation in the fermion picture generates a topologically non-trivial excitation 
by creating a domain wall (plus spin flip) in the spin picture, which then decays like a domain wall of TB fermions with time scale $vt$. In the paramagnetic phase the same excitation would
create a local excitation also in the spin picture, i.e. no domain wall.

Other observables, however, are different between the TFI model and TB fermions.
The transverse magnetization $\braket{\S^z}$ is finite in the TFI model 
(see \ref{sec:TFI_analytics})
while the corresponding quantity $\langle c^\dagger +c \rangle$ vanishes for TB fermions.
The bipartite entanglement $S_{\rm ent}(n,t)$ in the TFI model also develops a step structure,
but it is at all times smaller than for TB fermions (see \ref{sec:unscaled_results}) and it exhibits different scaling behaviour (see \Fig{fig:TFI_combined}). 
This fact only becomes fully apparent at large enough times, which our approach can provide.
It would be interesting if the above identity between TB fermions and the TFI model could be understood in more detail analytically.

\subsection{XXZ model}
\label{sec:XXZ_results}
\begin{figure}[tb]
\centerline{\includegraphics[width=\textwidth]{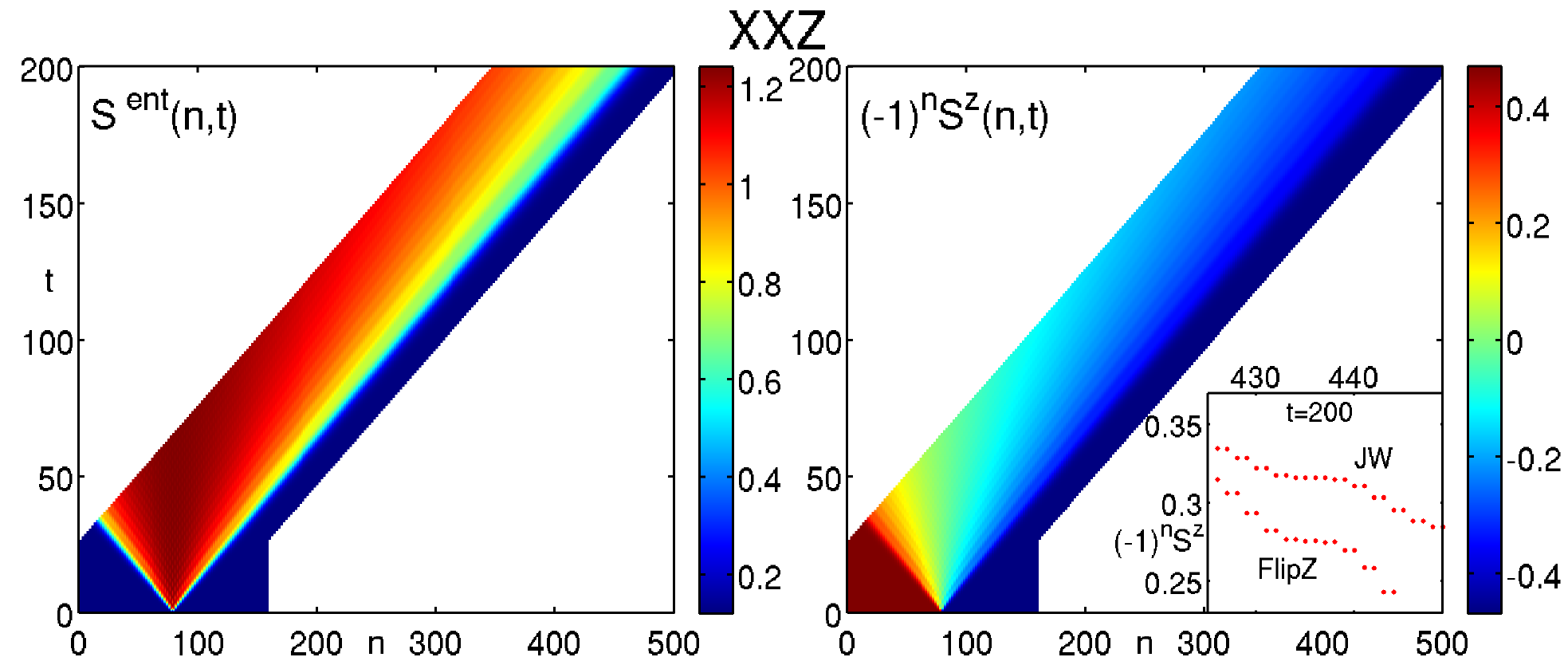}}
\label{XXZ_old}
\caption{
Time evolution of bipartite entanglement entropy $S_{\rm ent}(n,t)$ and staggered magnetization $\tilde{S}^z(n,t)$ 
in the XXZ antiferromagnet at $\Delta=-4$ after a JW excitation. We show times up to $t=200$ in order to keep structures resolvable to the eye, while simulations were performed up to $t=1000$.
\textit{Inset:} Magnification of the signal front at $t=200$ showing an internal step structure due to pairing.
}
\label{fig:XXZ_TimePlot}
\end{figure}

\begin{figure}[tb]
 \centering
 \includegraphics[width=0.9\linewidth,keepaspectratio=true]{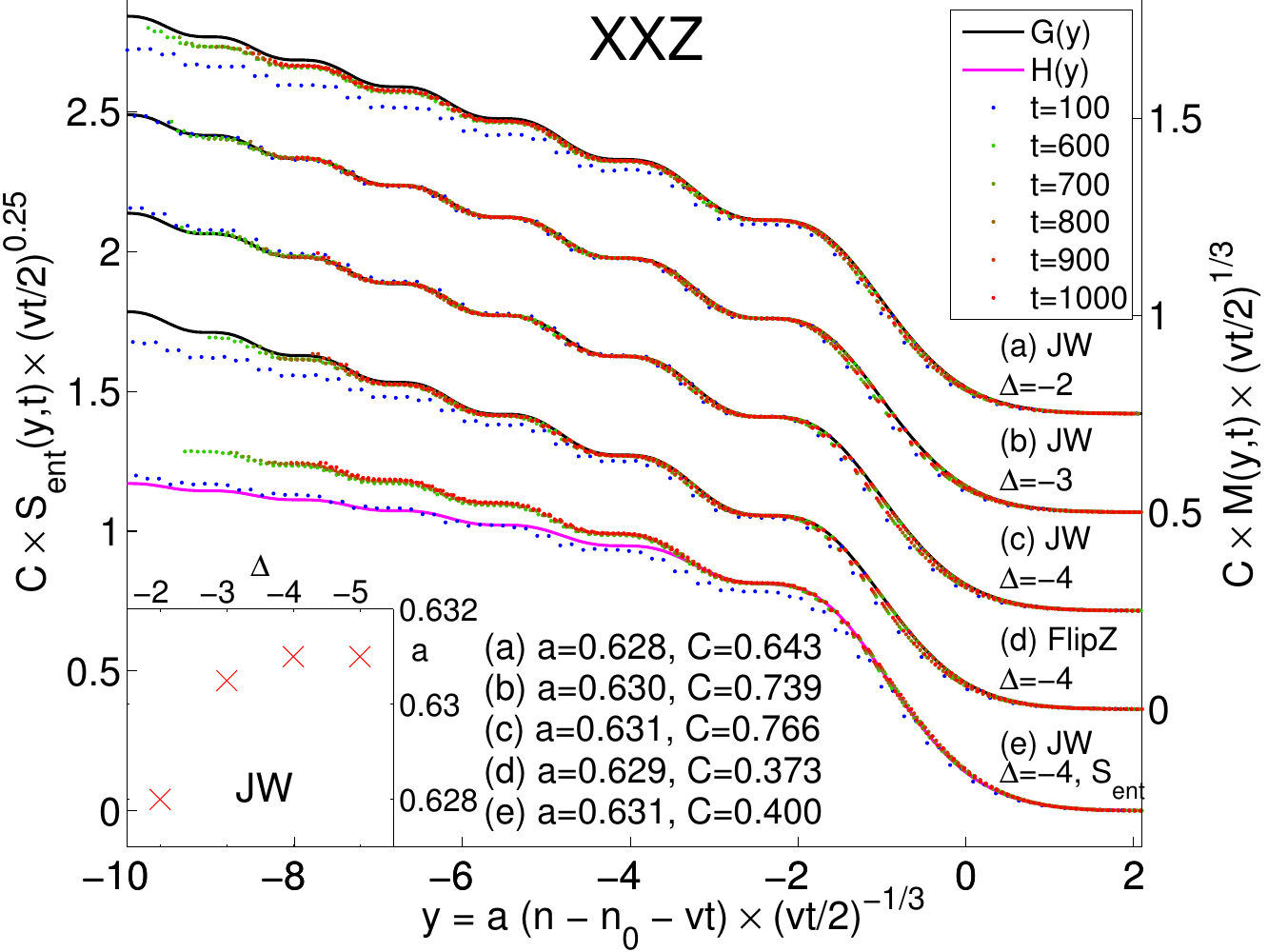}
 \caption{
\textit{Scaled} staggered normalized excess magnetization $M(y,t)$
(right axis) and bipartite entanglement entropy $S_{\rm ent}(y,t)$
(bottom only, left axis) vs. scaled position $y$ at the signal front for the XXZ model at different $\Delta$ for various signal types.
$G(y)$ and $H(y)$ are the same scaling functions as in \Fig{fig:TFI_combined}.
The lines are successively offset by $0.25$ in vertical direction.
\textit{Inset:} Horizontal scaling parameter $a$ as a function of $\Delta$ for JW excitations.
}
 \label{fig:XXZ_combined}
\end{figure}

Inspired by the above observations in the TFI model in the symmetry broken ferromagnetic phase, we also investigate the XXZ antiferromagnet
\cite{Barmettler0910,Fagotti13,Fagotti14,Pollmann13,Brockmann14,Gobert05,Pereira08,Langer0911,Lancaster10,Mossel10,Foster11,Jesenko11,Ganahl12,Alba14,Karrasch14,Sabetta13,Halimeh14},
\begin{equation}
\hat{H}=- \sum_n ( \hat{S}^x_n \hat{S}^x_{n+1} + \hat{S}^y_n \hat{S}^y_{n+1} + \Delta \hat{S}^z_n \hat{S}^z_{n+1} ),
\label{eq:XXZ_ham}
\end{equation} 
in the gapped symmetry broken phase for several $\Delta<-1$, where the ground state is also twofold degenerate.
We prepare the system in the maximally symmetry broken ground state $\ket{\Downarrow}$ 
with staggered magnetization $\tilde{S}^{z}_{\rm GS}=(-1)^n \braket{\S^z_n}<0$ using iDMRG
and again study the evolution of a JW excitation 
\begin{equation}
( c^{\dagger}_{n_{0}} + c_{n_{0}} )\ket{\Downarrow}=\prod_{n<n_{0}}( -2\S^{x}_{n} )( 2\S^{z}_{n_{0}} )\ket{\Downarrow}
\label{eq:JW_excitation_XXZ}
\end{equation} 
at site $n_{0}$ inside the window
(\Fig{fig:XXZ_TimePlot}).

Notice that due to $S^{x}_{\rm GS}=0$ a JW excitation is locally indistinguishable from a simple domain wall according to the magnetization and that
the roles of $x$ and $z$ are interchanged with respect to TFI results. 
Additionally, we also study a spin flip in $z$-direction at site $n_{0}$ (FlipZ). Window movement is triggered by bipartite entanglement entropy, resulting in window velocities consistent with exact results (see \ref{sec:XXZ_analytics}). During time evolution we use a second order Suzuki-Trotter decomposition with a step size of $\tau=0.01$ and maximum matrix dimensions of
$m_{\rm max}=150$ for $\Delta=-4$, $m_{\rm max}=160$ for $\Delta=-3$ and $m_{\rm max}=180$ for $\Delta=-2$ with discarded weights of at most $\Or(10^{-8})$.

The signal front again develops a step structure.
To our knowledge this had not been realized before our study, however it was recently confirmed \cite{Alba14,Halimeh14} after the preprint version of our study, but not further investigated.
We also observe a pairing effect between neighbouring spins, leading to an additional internal step structure, which stems from the spinon like nature of elementary excitations created by the quench (\Fig{fig:XXZ_TimePlot} inset).
Due to the dynamics generated by \eref{eq:XXZ_ham}, elementary spinons can only hop by two lattice sites at a time.

We find that at very large times -- which are virtually impossible to access with conventional MPS techniques \cite{Gobert05,Jesenko11} -- the staggered normalized excess magnetization 
\begin{equation}
M(n,t):=[\tilde{S}^{z}(n,t)-\tilde{S}^{z}_{\rm GS}]/|2\tilde{S}^{z}_{\rm GS}|
\label{eq:stag_excess_mag}
\end{equation} 
at the signal front
shows the same scaling behaviour as TB fermions, albeit with an additional horizontal scaling constant $a$, which is parameter dependent and increases with $|\Delta|$ (\Fig{fig:XXZ_combined} and inset). 
We therefore again interpret magnetization steps as due to individual propagating quasi-particles, which however show interaction effects by getting squeezed together more and more around the signal front with increasing $|\Delta|$. This behaviour can be explained by the fact that particles repel each other more with increasing interaction, but at the same time they are confined within the light cone dictated by the Lieb-Robinson bound. Since the particle density is much lower around the signal front, more and more particles are pushed towards the signal front and get squeezed together there. Our data however suggests that this effect saturates around $|\Delta|\approx 5$ (see inset of \Fig{fig:XXZ_combined}). It would be very interesting to understand these interaction effects between individual steps in more detail analytically.

The asymptotic scaling function $G(y)$ is approached differently for different $\Delta$, but the scaling exponents appear to be independent of $\Delta$ for all quenches investigated.
For $M(n,t)$ they are equal to the TB case with value $1/3$, whereas we again find a different effective exponent of $\approx1/4$ for the bipartite entanglement entropy (\Fig{fig:XXZ_combined}). 

\section{Conclusions}

We have introduced an easy-to-implement method combining finite and infinite system MPS techniques that can follow the propagation of a signal front 
on an infinite spin chain unimpeded and free from finite size effects for very long simulation times
and with very high precision, considerably improved over other approaches.
We note that even when the window is not moved, local signals can be simulated on the background of an infinite system,
without perturbations emanating from the boundary.
In this scenario the signal can be temporarily absorbed by the boundary,
though it is always reflected eventually.

Furthermore, the method is not restricted to the evolution of excitations under uniform Hamiltonians.
For example, the AKLT model \cite{AKLT} with inhomogeneous bond interactions
or 1D quantum systems under exponential or hyperbolic deformation~\cite{exp,cosh}
have uniform ground states, whereas the Hamiltonians are not uniform. 

To simulate the time evolution of a signal front of width $L$ propagating with velocity $v$ up to some time $t$, our method requires numerical effort of the order $\Or(Lt)$, whereas for
the same calculation using standard finite size MPS techniques numerical effort 
would scale as $\Or(Lt + vt^{2})$, i.e. with an additional $v$-dependent factor which scales 
quadratically in simulation time. We want to emphasize that 
additionally, standard finite size MPS techniques would also suffer from finite size effects such as boundary effects or the absence of exact ground state degeneracies in symmetry broken phases.

We have found that for all local quenches investigated in the symmetry broken phases of the TFI and the XXZ model,
distinct magnetization plateaus develop at the emerging signal front  at very large times,
where the scaled excess magnetizations in both models show the same long time limit scaling behaviour
as the particle density of TB fermions after an initial domain wall excitation. For TB fermions these
plateaus have recently been understood as due to individual propagating particles \cite{Eisler13}. 
Because of their quantum origin these plateaus can not be studied \cite{Eisler13,RiegerComm} by means of semiclassical approaches such as in \cite{Rieger11}.

 Our method has enabled us to calculate the time evolution of the order parameters of both models around the signal fronts generated by local quenches and investigate their features, which to our knowledge are available neither analytically nor semi-classically. In all cases it is important to reach very large simulation times -- which are 
easily accessible through our approach -- in order to reach the proper scaling regimes.

In the XXZ model we have observed an additional internal step structure due to the spinon nature of the involved elementary excitations, as well as
parameter dependent interaction effects between individual plateaus in the form of increasing spatial compression of the plateau width close 
to the signal front. This effect appears to saturate for $|\Delta|\gg 1$.
For the TFI model we have additionally found a surprising exact agreement of the normalized excess longitudinal magnetization after a JW excitation with the density of TB fermions after a domain wall excitation. This exact mapping however does not apply to other observables such as e.g. bipartite entanglement.

It would be interesting to understand both the interaction effects between plateaus in the XXZ model and the exact agreement between the TFI model and TB fermions in more detail analytically.

\ack
We would like to thank Th.~Barthel, V.~Eisler, F.~Maislinger, M.M.~Rams, D.~Schuricht, U.~Schollw\"ock, and F.~Verstraete for valuable discussions.  
This work was supported by the Austrian Science Fund (FWF): F4104 SFB ViCoM and by the EP - SRC under grant
EP/I032487/1. T.N. acknowledges the support of Grant-in-Aid for Scientific Research (C) No. 22540388.

\newpage	
\appendix
\setcounter{section}{0}

\section{CMW Time Evolution and Boundary Update Methods}
\label{sec:timeevo}

In this appendix we illustrate one time evolution step for the entire
system when following a right moving signal. We describe the procedure in the following
order. We first evolve the part of the system contained within the CMW (\ref{sec:CMW_update}) before
updating the right part using Method I (\ref{sec:method_I}) and updating the left part using 
the more involved 
Method II
(\ref{sec:method_II} and \ref{sec:ren_ham}). Note that this is the setup used in the main text, however in principle any of the two methods can be used at any boundary. A detailed assessment of different setups is given in \ref{sec:precision}. We also describe
the process of moving the CMW along with a propagating signal (\ref{sec:window_movement}). 
A short sketch of both boundary update methods, illustrating their advantages and restrictions, along with a motivation of the above choice is given in the main text.

\subsection{System Initialization}
\label{sec:init}
In the main text in particular we use a setup dividing the system into a semi-infinite, initially translation invariant left part, a finite-size CMW (inside of which a signal will be created) 
and a semi-infinite, at all times translation invariant right part. We initialize the system by first determining a uniform MPS representation of the respective model's ground state on an infinite 
chain using iDMRG \cite{idmrg1,idmrg2}. We then set all MPS matrices inside the CMW (matrices $A^{\sigma_{j}}$ and $B^{\sigma_{j}}$), the semi-infinite right part (matrices $R_{A}^{\sigma}$ and
$R_{B}^{\sigma}$ forming this part's two-site unit cell) and the semi-infinite left part (\textit{all} matrices $L^{\sigma_{j}}$) to this uniform MPS ground state representation after appropriate 
(left or right) orthonormalization \cite{MPS2,idmrg2}, i.e. we initialize the entire system to be in the infinite system's translation invariant ground state. Subsequently we locally 
excite the system out of its ground state to generate several different kinds of local signals by applying suitable operators to one or more MPS matrices inside the CMW.

For other purposes the generalization to different initial conditions is straight forward.

\subsection{Time Evolution within the CMW (CMW Update)}
\label{sec:CMW_update}

Without loss of generality we consider a CMW with an even number of sites and first order even-odd
Suzuki-Trotter decomposition~\cite{TrotterSuzuki} with local operators $\hat{u}_{j,j+1}( \tau ) = \rme^{ -\rmi \tau {\hat
h}_{j,j+1}^{~} }$ and finite time steps $\tau$. The generalization to higher order Suzuki-Trotter decompositions and
windows containing an odd number of sites is straight forward. All simulations in this work have been performed using
second order Suzuki-Trotter decomposition and windows with an even number of sites.

For one time step inside the CMW we use tDMRG \cite{tdmrg} and apply
$\hat{u}_{j,j+1}( \tau ) = \rme^{ -\rmi \tau {\hat h}_{j,j+1}^{~} }$ to the bonds from
$(\ell+1,\ell+2)$ until $(r-1,r)$ and update matrices $A^{s_{j}}$ and $B^{s_{j}}$ contained within the CMW. The
junction bonds $(\ell,\ell+1)$ and $(r,r+1)$ at the left and right boundary of the CMW are updated separately. 

We first update all \textit{odd} bonds $\{\ldots,(r-3,r-2),(r-1,r)\}$ and then all \textit{even} bonds
$\{\ldots,(r-4,r-3),(r-2,r-1)\}$. The junction bonds $(\ell,\ell+1)$ and $(r,r+1)$ are thus defined to be even bonds
(see \Fig{fig:cmw_update}). By choosing this order we preserve the structure of the Suzuki-Trotter decomposition
of the CMW and the right part, when Method I is used to update the right boundary.

At this stage all even and odd bonds have been updated, except for the junction bonds $(\ell,\ell+1)$ and  $(r,r+1)$, 
i.e. the boundary matrices $A^{s_{\ell+1}}$ and $B^{s_{r}}$ are not yet fully updated.

\begin{figure}[t]
 \centering
 \includegraphics[width=\linewidth,keepaspectratio=true]{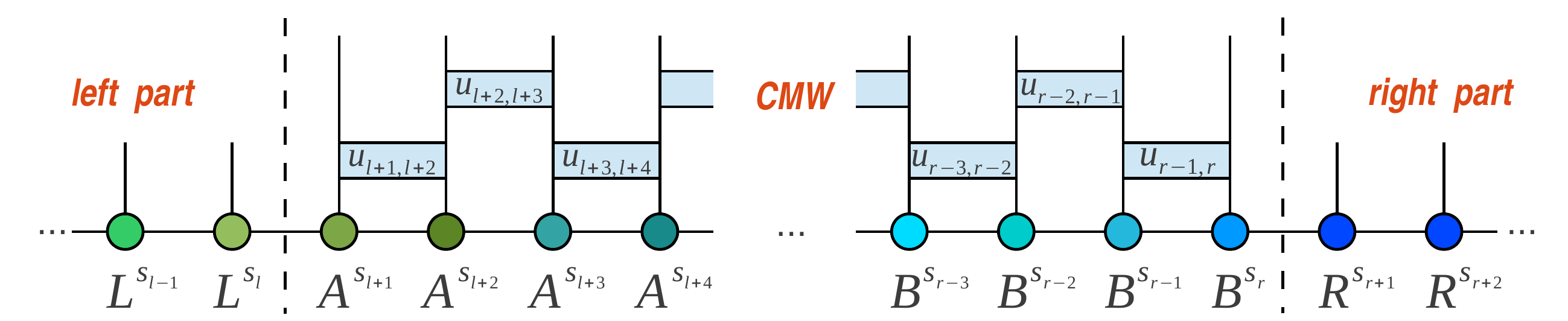}
 \caption{One time step of the CMW update in the case of first order Suzuki-Trotter decomposition and a CMW with an even
number of sites.}
 \label{fig:cmw_update}
\end{figure}

Note that an implementation of this update using Time Evolving Block Decimation (TEBD \cite{tebd}) is equivalent.
For a graphical representation see \Fig{fig:cmw_update}.

\subsection{Method I (Uniform Update).} 
\label{sec:method_I}
We use this easy to implement procedure for the right boundary. Due to the assumed translation invariance over a 2-site unit cell,
this part can be described by two right-orthogonal matrices
$R_{A}^{s_{j}}$ and $R_{B}^{s_{j}}$, such that the wavefunction in MPS representation around the right boundary reads
\begin{equation}
 \ldots B^{s_{r-1}} B^{s_r}
  R_{A}^{s_{r+1}} R_{B}^{s_{r+2}} R_{A}^{s_{r+3}} R_{B}^{s_{r+4}}\ldots.
\end{equation} 
The evolution of the matrices $R_{A}^{s_{j}}$ and $R_{B}^{s_{j}}$
is performed by iTEBD (or variations thereof) using local operators $\hat{u}^{A}(\tau)$ and $\hat{u}^{B}( \tau )$
~\cite{iTEBD,HastingsiTEBD}, where $\hat{u}^{A}(\tau)$ acts on \textit{odd} bonds and $\hat{u}^{B}( \tau )$ acts on
\textit{even} bonds. 

In a first step, we apply an \textit{odd} bond iTEBD update in the right part to get
\begin{equation}
R_{A\circ}^{s_{A}}R_{B\circ}^{s_{B}}=\sum_{s'_{A}s'_{B}}u(\tau)^{A}_{(s_{A}s_{B})(s'_{A}s'_{B})}R_{A}^{s'_{A}}R_{B}^{s'_
{B}},
\label{eq:itebdodd}
\end{equation}
where ${\circ}$ denotes matrices having received an odd bond update.
Here the decomposition of the result of the right hand side of
\eref{eq:itebdodd} is implicitly assumed. It can be done by an SVD
either involving a division by Schmidt values following \cite{iTEBD} or avoiding the division by Schmidt values by using the approach of \cite{HastingsiTEBD}.

The wavefunction 
at this point reads
\begin{equation}
 \ldots B^{s_{r-2}}_{\bullet} B^{s_{r-1}}_{\bullet} B^{s_r}_{\circ}
  R_{A\circ}^{s_{r+1}} R_{B\circ}^{s_{r+2}} \ldots,
\end{equation} 
where $\bullet$ denotes matrices having received both odd and even updates.

Special attention has to be paid to the operation of $\hat{u}_{r, r+1}(\tau)$ at the junction 
bond in order to update $B^{s_r}_{\circ}$. 
For this we form $\Phi^{s_{r}s_{r+1}}_{\circ\circ}:=B^{s_r}_{\circ}
R_{A\circ}^{s_{r+1}}$ and act with $\hat{u}_{r,r+1}$ to get $\Phi^{s_{r}s_{r+1}}_{\bullet\bullet}$.
In parallel we perform an \textit{even} bond iTEBD update in the right part 
to get 
\begin{equation}
 R_{B\bullet}^{s_{B}}R_{A\bullet}^{s_{A}}
=\sum_{s'_{A}s'_{B}}u(\tau)^{B}_{(s_{B}s_{A})(s'_{B}s'_{A})}R_{B\circ}^{s'_{B}}R_{A\circ}^{s'_{A}}.
\label{eq:itebdeven}
\end{equation} 
where again the decomposition of the result of the right side is implicitly assumed. 

All bonds have now been updated. Since there is negligible influence of the signal around the right boundary by
construction, the state of the right part should be the same as for a time evolved uniform system without signal up to
high precision, i.e. we can also assume $\Phi^{s_{r}s_{r+1}}_{\bullet\bullet}=B^{s_r}_{\bullet}R_{A\bullet}^{s_{r+1}}$,
where both $B^{s_r}_{\bullet}$ and $R_{A\bullet}^{s_{r+1}}$ are right-orthogonal and $R^{s_{r+1}}_{A\bullet}$ is
obtained from
\eref{eq:itebdeven}.
We extract $B^{s_r}_{\bullet}$ from  $\Phi^{s_{r}s_{r+1}}_{\bullet\bullet}$
by 
exploiting the right-orthogonality of $R_{A\bullet}^{s_{r+1}}$:
\begin{equation}
 B^{s_r}_{\bullet}=\sum_{s_{r+1}}\Phi^{s_{r}s_{r+1}}_{\bullet\bullet}R_{A\bullet}^{s_{r+1}\dagger}.
 \label{eq:B_from_Phi}
\end{equation} 
The wavefunction in MPS form is now completely updated around the right boundary and reads
\begin{equation}
 \ldots B^{s_{r-2}}_{\bullet} B^{s_{r-1}}_{\bullet} B^{s_r}_{\bullet}
  R_{A\bullet}^{s_{r+1}} R_{B\bullet}^{s_{r+2}} \ldots.
\end{equation} 

\begin{table}[t]
 \centering
 \fbox{
 \parbox{0.9\linewidth}
 {
 \begin{enumerate}
  \item Apply odd iTEBD update to get $R_{A\circ}^{s_{A}}$, $R_{B\circ}^{s_{B}}$.
  \item Use $R_{A\circ}^{s_{A}}$ to form $\Phi^{s_{r}s_{r+1}}_{\circ\circ}=B^{s_r}_{\circ}
 R_{A\circ}^{s_{r+1}}$ 
  \item Apply $\hat{u}_{r,r+1}$ to get $\Phi^{s_{r}s_{r+1}}_{\bullet\bullet}$.
  \item Apply even iTEBD update to get $R_{B\bullet}^{s_{B}}$, $R_{A\bullet}^{s_{A}}$.
  \item Use $R_{A\bullet}^{s_{A}}$  to obtain
$B^{s_r}_{\bullet}=\sum_{s_{r+1}}\Phi^{s_{r}s_{r+1}}_{\bullet\bullet}R_{A\bullet}^{s_{r+1}\dagger}$.
 \end{enumerate}
 }
 }
 \caption{Algorithm for \textit{Method I} updating the right boundary of the CMW. For a graphical representation see
 \Fig{fig:itebd_update}.}
 \label{tab:itebdalg}
\end{table}


\begin{figure}[t]
 \centering
 \includegraphics[width=0.9\linewidth,keepaspectratio=true]{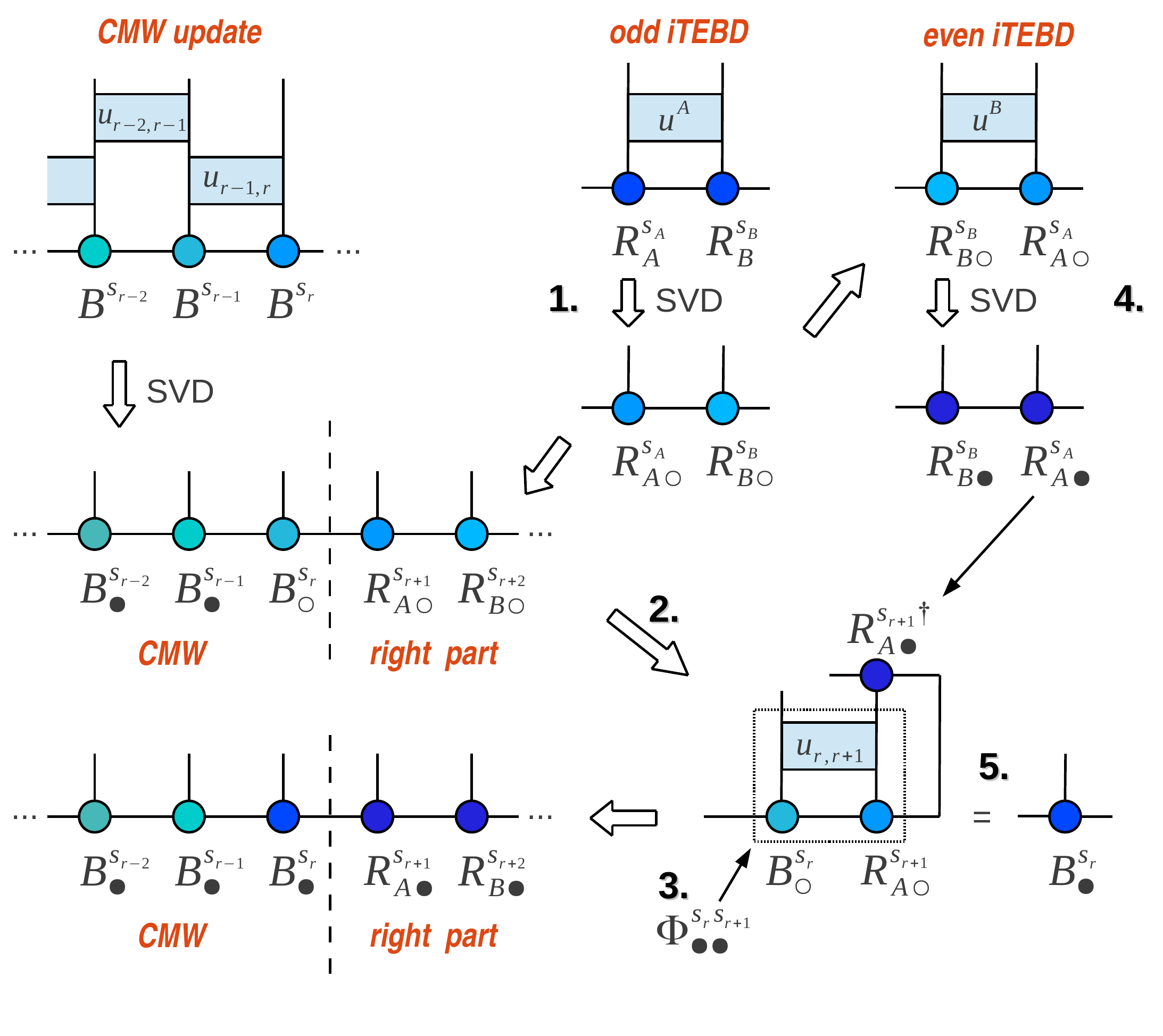}
 \caption{Graphical representation for updating the right boundary of the CMW with Method I according to the steps in
table \ref{tab:itebdalg}.}
 \label{fig:itebd_update}
\end{figure}


We note that in general, the decomposition $\Phi^{s_{r}s_{r+1}}_{\bullet\bullet}=\tilde{B}^{s_r}_{\bullet}
\tilde{R}_{\bullet}^{s_{r+1}}$ into right-orthogonal matrices is not unique, but involves a gauge freedom
$\tilde{B}^{s_r}_{\bullet} \tilde{R}_{\bullet}^{s_{r+1}}=B^{s_r}_{\bullet}x^{-1}x R_{A\bullet}^{s_{r+1}}$ with $x$ a
unitary matrix 
(Exploiting right orthogonality of both $\tilde{R}_{\bullet}^{s_{r+1}}$ and
$R_{A\bullet}^{s_{r+1}}$ we have $\unity=\sum_{s}\tilde{R}_{\bullet}^{s_{r+1}}\tilde{R}_{\bullet}^{s_{r+1}\dagger}=
x\sum_{s}R_{A\bullet}^{s_{r+1}}R_{A\bullet}^{s_{r+1}\dagger}x^{\dagger}=xx^{\dagger}$. 
Since $x$ is square this also means $x^{\dagger}x=\unity$ and thus $x$ is unitary.).

If the decomposition $\Phi^{s_{r}s_{r+1}}_{\bullet\bullet}=\tilde{B}^{s_r}_{\bullet}
\tilde{R}_{\bullet}^{s_{r+1}}$ was carried out in the standard TEBD/tDMRG way (i.e. by means of an SVD), then a
different gauge
$\tilde{R}_{\bullet}^{s_{r+1}}\neq R_{A\bullet}^{s_{r+1}}$ and thus $\tilde{B}^{s_r}_{\bullet}\neq B^{s_r}_{\bullet}$
 would result in general, since $\tilde{R}_{\bullet}^{s_{r+1}}$ was produced
algorithmically in a different way than $R_{A\bullet}^{s_{r+1}}$. 
In that case, i.e. if $\tilde{B}^{s_r}_{\bullet}$ was used instead of $B^{s_r}_{\bullet}$, incompatible basis sets would
meet at the junction bond, which would result in perturbations spreading 
from the boundary. By use of \eref{eq:B_from_Phi} we ensure that the correct gauge is chosen automatically.

This concludes one time step for the right part and right boundary. For an algorithmic summary see table
\ref{tab:itebdalg}, for a detailed graphical representation see \Fig{fig:itebd_update}. 

The procedure can also be easily translated to the left boundary exploiting
left orthogonality, where translation invariance of the left-orthogonal matrices $L^{s_{j}}$ is then required. 

Method I is also applicable when $H_{R}$ is time dependent, e.g. in case of a global quench.

\subsection{Method II (Renormalized Update)}
\label{sec:method_II}

We use this procedure for the left boundary. For this Method we follow a similar approach as introduced by
Cazalilla and Marston \cite{non_adapt} (Method II is similar to the algorithm introduced in \cite{phien-infinite}, where 
preprints of \cite{phien-infinite} and of the present paper appeared at the same time),
such that matrices $L^{s_{j}}$ in the left part remain unchanged at all times during time evolution.

The effect of the left part is encoded in a renormalized formulation of $\hat{H}_{\rm
\triangleleft,\ell+1}:=\hat{H}_{\rm L} + \hat{h}_{\ell,\ell+1}$, which is exactly the renormalized
Hamiltonian used in standard DMRG formulations (see e.g. \cite{MPS1,MPS2}). All changes in the left part are then solely
encoded in an update of the boundary matrix $A^{s_{\ell+1}}$.
Note that for this method the matrices $L^{s_{j}}$ in the left part need not be translation invariant.

Since matrices $L^{s_{j}}$ are not changed during this update, we rewrite the wavefunction in MPS form
after the CMW update in terms of the auxiliary basis states
$\ket{a_{\ell}}=\sum_{s_{j\leq\ell}}\left(\ldots L^{s_{\ell-1}}L^{s_{\ell}}\right)_{a_{\ell}}\ket{\ldots
s_{\ell-1}s_{\ell}}$
connecting $L^{s_{\ell}}$ and $A^{s_{\ell+1}}$:
\begin{equation}
\Psi(a_{\ell}\vert s_{\ell+1}^{\circ}s_{\ell+2}^{\bullet}\ldots;t) 
=\left(A_{\circ}^{s_{\ell+1}}A_{\bullet}^{s_{\ell+2}} \ldots\right)_{a_{\ell}},
\label{eq:psirenorm}
\end{equation} 
where $a_{\ell}$ is the left index of matrix $A_{\circ}^{s_{\ell+1}}$. 
The right hand side of \eref{eq:psirenorm} is formally just the semi-infinite product of all matrices to the right of
site $\ell$.
The overall state vector after the CMW update can thus also be written
\begin{equation}
 \ket{\Psi(t)}=\sum_{a_{\ell}}\sum_{s_{\ell+1}\ldots}\Psi(a_{\ell}\vert
s_{\ell+1}^{\circ}s_{\ell+2}^{\bullet}\ldots;t) 
\ket{a_{\ell}}\ket{s_{\ell+1}\ldots}.
\end{equation} 
Here $\bullet$ and $\circ$ again mark sites which have received a complete and incomplete update respectively. Notice
that in Method II the basis $\{\ket{a_{\ell}}\}$ will remain unchanged at all times. For a graphical representation of
\eref{eq:psirenorm} see \Fig{fig:Heff_misc}(a).

\begin{table}[t]
 \centering
 \fbox{
 \parbox{0.9\linewidth}
 {
 \begin{enumerate}
  \item Perform only once for each CMW position:
  \begin{enumerate}
  \item Determine renormalized expression $H^{\rm eff}_{\rm \triangleleft,\ell+1}$ of $\hat{H}_{\rm
\triangleleft,\ell+1}$, formulated in
block-spin basis $\{\ket{a_{\ell}s_{\ell+1}}\}$.
  \item Calculate $U\left(\tau\right)^{\rm eff}_{\rm \triangleleft,\ell+1}=\exp\left(-\rmi\tau H^{\rm eff}_{\rm
\triangleleft,\ell+1}\right)$.
  \end{enumerate}
  \item Update $A_{\circ}^{s'_{\ell+1}}$ using $U^{\rm eff}_{\rm \triangleleft,\ell+1}$ in each time step to get
$A_{\bullet}^{s_{\ell+1}}=
 \sum_{s'_{\ell+1}} U(\tau)_{\rm \triangleleft,\ell+1}^{{\rm eff},s_{\ell+1}s'_{\ell+1}}A_{\circ}^{s'_{\ell+1}}$.
 \end{enumerate}
 }
 }
 \caption{Algorithm for \textit{Method II} updating the left boundary of the CMW. For a graphical representation see
\Fig{fig:Heff_update}.}
 \label{tab:Heffalg}
\end{table}

We now need a renormalized representation $H^{\rm eff}_{\rm \triangleleft,\ell+1}$ of $\hat{H}_{\rm
\triangleleft,\ell+1}$
formulated in the
block-spin basis $\{\ket{a_{\ell}s_{\ell+1}}\}$. A possible method to calculate $H^{\rm eff}_{\rm
\triangleleft,\ell+1}$ is outlined in \ref{sec:ren_ham}.
Once we have such a renormalized expression we can determine the renormalized time evolution operator for the left part
\begin{equation}
 U\left(\tau\right)^{\rm eff}_{\rm \triangleleft,\ell+1}:=\exp\left(-\rmi\tau H^{\rm eff}_{\rm
\triangleleft,\ell+1}\right),
\label{eq:Ueff}
\end{equation} 
where we use the same small time step $\tau$ as for the Suzuki-Trotter updates. This time evolution operator is then
used to update $\Psi(a_{\ell}\vert s_{\ell+1}^{\circ}s_{\ell+2}^{\bullet}\ldots ; t)$. However, as it is a unitary
operator defined in the block-spin basis $\{\ket{a_{\ell}s_{\ell+1}}\}$, it only updates $A_{\circ}^{s_{\ell+1}}$ and we
get
\begin{equation}
 A_{\bullet}^{s_{\ell+1}}=
 \sum_{s'_{\ell+1}} U(\tau)_{\rm \triangleleft,\ell+1}^{{\rm eff},s_{\ell+1}s'_{\ell+1}}A_{\circ}^{s'_{\ell+1}}.
 \label{eq:Heffupdate}
\end{equation} 

\begin{figure}[t]
 \centering
 \includegraphics[width=0.9\linewidth,keepaspectratio=true]{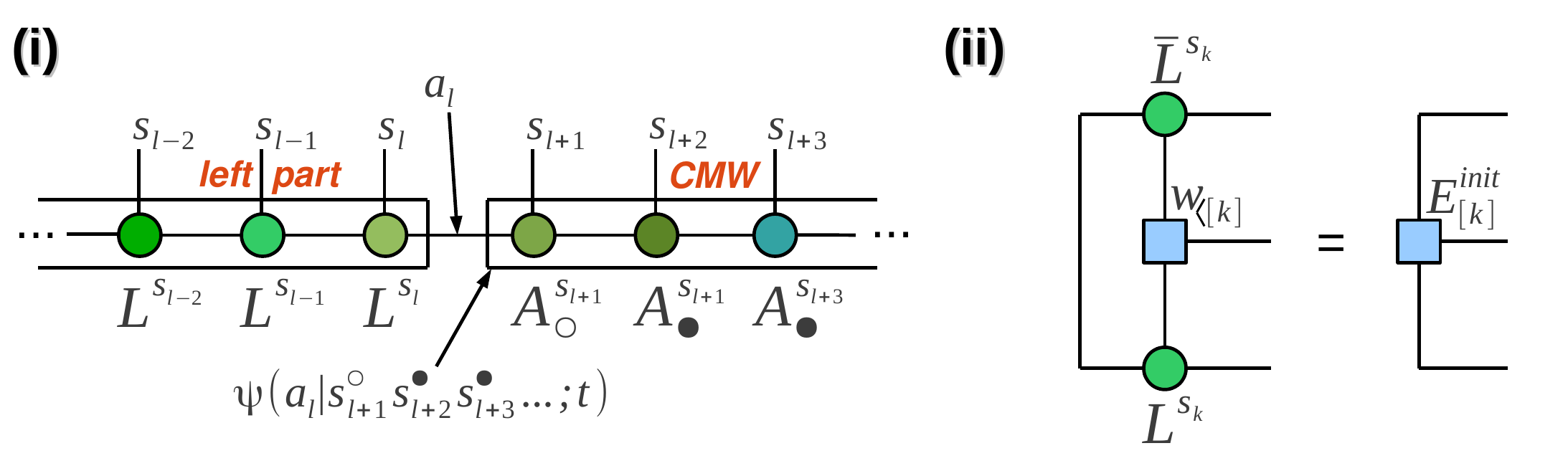}
\caption{(i) Graphical Representations of the definition of $\Psi(a_{\ell}\vert
s_{\ell+1}^{\circ}s_{\ell+2}^{\bullet}\ldots;t)$ in \eref{eq:psirenorm}. Note that the matrices $L^{s_{j}}$ within the
left part and the basis $\{\ket{a_{\ell}}\}$ remain unchanged during the simulation for Method II. (ii) Graphical
representation of the construction of the initial element
$E^{\rm init}_{[k]}$ defined in \eref{eq:Einitial} to approximate the semi-infinite product $E_{[\ell]}$.}
 \label{fig:Heff_misc}
\end{figure}

\begin{figure}[t]
 \centering
 \includegraphics[width=0.9\linewidth,keepaspectratio=true]{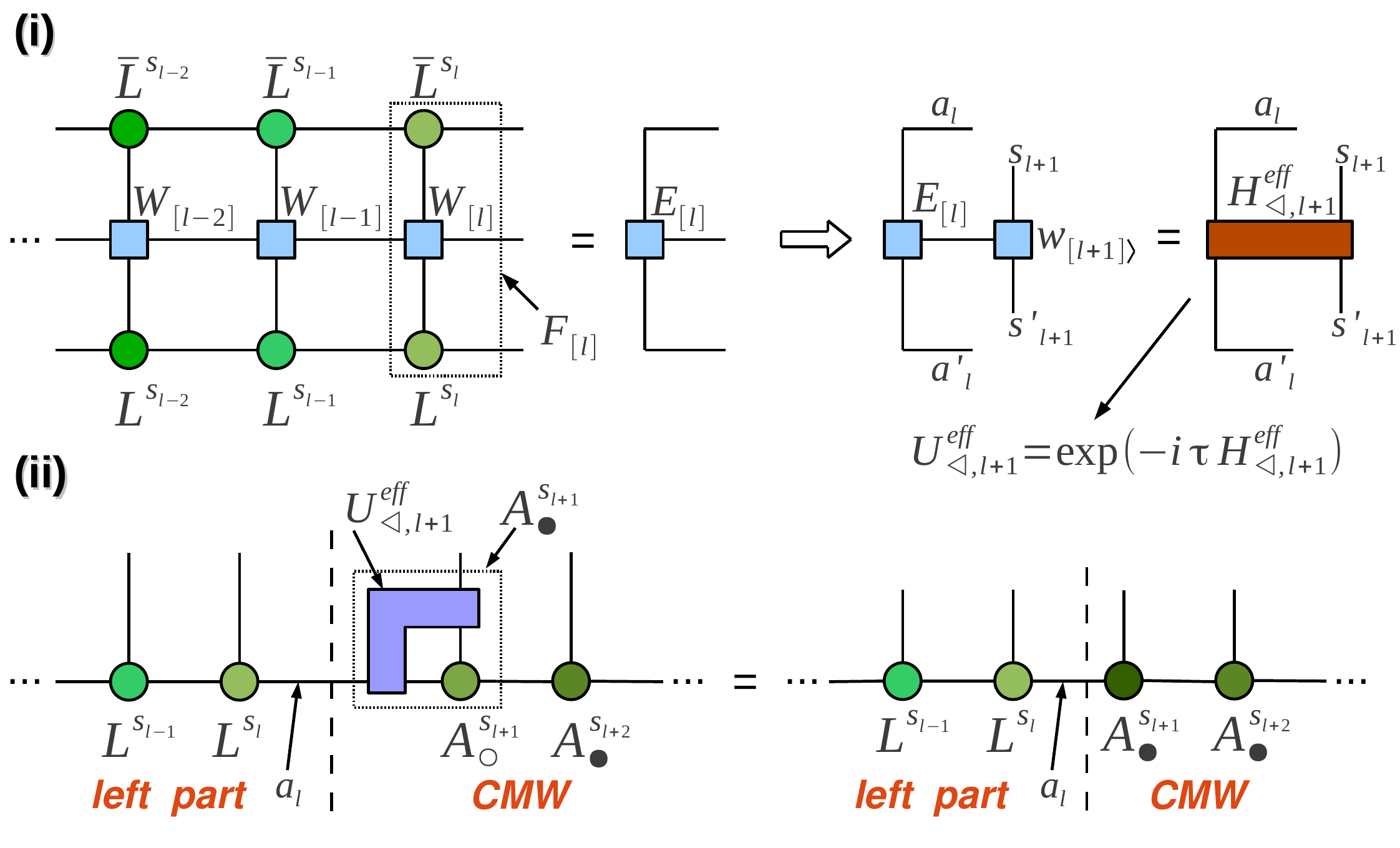}
 \caption{Graphical representation for updating the left boundary of
   the CMW with Method II with the steps listed in
table \ref{tab:Heffalg}. (i) Constructing the renormalized Hamiltonian $H^{\rm eff}_{\rm
\triangleleft,\ell+1}$ as outlined in \ref{sec:ren_ham} and there defined in \eref{eq:Heff}. (ii) Using the
renormalized time evolution operator $U(\tau)^{\rm
eff}_{\rm \triangleleft,\ell+1}$ as defined in \eref{eq:Ueff} to update the left boundary matrix $A^{s_{\ell+1}}$
according to \eref{eq:Heffupdate}.
}
 \label{fig:Heff_update}
\end{figure}

This concludes one time step for the left boundary. Notice that the matrices in the left part are \textit{not} updated
as all change in the left part are compressed into the boundary matrix
$A^{s_{\ell+1}}$ with constant basis $\{\ket{a_{\ell}}\}$. In this
sense the update is non-adaptive. 

Notice also that $U^{\rm eff}_{\rm \triangleleft,\ell+1}$ breaks the structure of the even-odd
Suzuki-Trotter decomposition in the left part. This introduces an additional error, which is of the same order as the
Suzuki-Trotter error and can in principle be made arbitrarily small by using higher order Suzuki-Trotter decompositions
and smaller time steps $\tau$ at the cost of increased computational time. The effect of this additional error is
investigated in detail in \ref{sec:precision}. It could be avoided by using the renormalized
imaginary-time transfer matrix, as used in finite temperature DMRG \cite{XiangWangShibata}, to update
$A_{\circ}^{s_{\ell+1}}$. For an algorithmic summary see table
\ref{tab:Heffalg}, for a graphical representation of this update see \Fig{fig:Heff_update}.

\subsection{Renormalized Hamiltonian for Method II}
\label{sec:ren_ham}

For determining $H^{\rm eff}_{\rm \triangleleft,\ell+1}$ used in Method II, we first assume a left part that is
semi-infinite. Consider $\hat{H}$ in MPO form \cite{mpo1,mpo2,mpo3,mpolong,mpoinf}
\begin{equation}
 \hat{H} = \sum_{\boldsymbol s} \prod_{j=-\infty}^{\infty}W_{[j]}^{s_{j}s'_{j}}\ket{\boldsymbol s}\bra{\boldsymbol s'},
\end{equation} 
where $W_{[j]}^{s_{j}s'_{j}}$ are matrices of some dimension $d_{W}\times d_{W}$ containing operator elements
$O^{s_{j}s'_{j}}$. This decomposition can also be written in operator form, where we define
$\hat{W}_{[j]}:=\sum_{s_{j}s'_{j}}W_{[j]}^{s_{j}s'_{j}}\ket{s_{j}}\bra{s'_{j}}$ as $d_{W}\times d_{W}$ matrices
containing operators. We can then simply write
$ \hat{H} = \prod_{j=-\infty}^{\infty}\hat{W}_{[j]}.$
For finite size (or semi-infinite) operators, this product of MPO matrices is terminated by $d_{W}$-dimensional
operator-valued boundary vectors $\hat{w}_{\langle[j]}$ and (or) $\hat{w}_{[j]\rangle}$.

An example for an MPO decomposition for the Transverse Field Ising (TFI) Hamiltonian $\hat{H}_{\rm
TFI}=-\sum_{j}\hat{S}^{x}_{j}\hat{S}^{x}_{j+1}-h\sum_{j}\hat{S}^{z}_{j}$ is given by
\begin{equation}
 \hat{W}_{[j]}=
 \left[\begin{array}{ccc}
\hat{\unity}_{j}&0&0\\
-\hat{S}^{x}_{j}&0&0\\
-h\hat{S}^{z}_{j}&\hat{S}^{x}_{j}&\hat{\unity}_{j}
\end{array}\right]
\end{equation} 

\begin{equation}
 \hat{w}_{\langle[j]}=
  \left[\begin{array}{ccc} -h\hat{S}^{z}_{j}& \hat{S}^{x}_{j}&\hat{\unity}_{j}\end{array}\right]\qquad
  \hat{w}_{[j]\rangle}=
  \left[\begin{array}{ccc}\hat{\unity}_{j} & -\hat{S}^{x}_{j}&-h\hat{S}^{z}_{j} \end{array}\right]^{T}.
\end{equation}

For the TFI Hamiltonian we thus have $d_{W}=3$.



We can express $\hat{H}_{\rm \triangleleft,\ell+1}$ in terms of an MPO as
\begin{equation}
 \hat{H}_{\rm \triangleleft,\ell+1}=\ldots \hat{W}_{[\ell-1]}\hat{W}_{[\ell]}\hat{w}_{[\ell+1]\rangle},
\end{equation} 
where we have terminated the semi-infinite product of MPO matrices with the boundary vector
$\hat{w}_{[\ell+1]\rangle}$.

In order to get $H^{\rm eff}_{\rm \triangleleft,\ell+1}$ we use matrices $L^{s_{j}}$ to renormalize $\hat{H}_{\rm
\triangleleft,\ell+1}$. For this, consider the $d_{W}\times d_{W}$ dimensional \textit{MPO transfer matrix} defined as
\begin{equation}
 F_{[j]}^{b_{j-1}b_{j}}:=\sum_{s_{j}s'_{j}}W_{[j],b_{j-1}b_{j}}^{s_{j}s'_{j}}\bar{L}^{s_{j}}\otimes L^{s'_{j}},
 \label{eq:mpo_transfer}
\end{equation} 
containing $m^2\times m^2$ matrices, where $m$ is the matrix dimension of the MPS matrices $L^{s_{j}}$ and
$\bar{L}^{s_{j}}$ denotes the complex conjugate of $L^{s_{j}}$. 

$H^{\rm eff}_{\rm \triangleleft,\ell+1}$ can then be written 
as
\begin{eqnarray}
 H^{{\rm eff},s_{\ell+1}s'_{\ell+1}}_{{\rm \triangleleft,\ell+1}}
 &=\sum_{b_{\ell}=1}^{d_{W}}\Big(\prod_{j\leq\ell}F_{[j]}\Big)_{b_{\ell}}w_{[\ell+1]\rangle,b_{\ell}}^{s_{\ell+1}
s'_{\ell+1}} \\
 &=\sum_{b_{\ell}=1}^{d_{W}}E_{[\ell]}^{b_{\ell}}w_{[\ell+1]\rangle,b_{\ell}}^{s_{\ell+1}s'_{\ell+1}},
 \label{eq:Heff}
\end{eqnarray} 
where we have defined the semi-infinite product 
\begin{equation}
E_{[\ell]}^{b_{\ell}}:=\Big(\prod_{j\leq\ell}F_{[j]}\Big)_{b_{\ell}}.
\label{eq:E}
\end{equation} 
$H^{{\rm eff},s_{\ell+1}s'_{\ell+1}}_{{\rm \triangleleft,\ell+1}}$ is then a set of $m\times m$ matrices labelled by
$s_{\ell+1}$ and $s'_{\ell+1}$ and $E_{[\ell]}$ can be understood as a $d_{W}$-dimensional vector containing $m\times m$
matrices. For a graphical representation of these steps see \Fig{fig:Heff_update}(1). Note that the vector element
$E_{[j]}^{1}$ accumulates the renormalized Hamiltonian containing all sites
$k\leq j$ (see e.g. \cite{mpo2}).
%

To determine \eref{eq:Heff} we need a way to calculate the semi-infinite matrix product $E_{[\ell]}$. 
For the moment we consider the case where both $\hat{H}_{L}$ and the matrices $L^{s_{j}}$ are translation
invariant.
In this case $F_{[j]}$ is also translation invariant and $E_{[\ell]}$ can be calculated by e.g. finding the dominant
left eigenvector of $F_{[j]}$ as explained in \cite{mpoinf}.

However here we follow an approximate but sufficiently accurate approach for calculating
$E_{[\ell]}$, which is inspired by standard DMRG formulations. For this we relax the condition of semi-infinity for
the left Hamiltonian $\hat{H}_{\rm \triangleleft,\ell+1}$ and approximate it with a finite size Hamiltonian, which we
increase in size until we get a converged result. The finite size version of $\hat{H}_{\rm \triangleleft,\ell+1}$ in MPO
form is thus contracted also on the left side by $\hat{w}_{\langle[k]}$ for some $k\ll\ell$. We first
compute an
initial $E_{[k]}$ as
\begin{equation}
 E^{\rm init}_{[k],b_{k}}=\sum_{s_{k}s'_{k}}w_{\langle[k],b_{k}}^{s_{k}s'_{k}}L^{s_{k}\dagger}L^{s'_{k}},
 \label{eq:Einitial}
\end{equation} 
exploiting the left-orthogonality of the matrices $L^{s_{j}}$. For a graphical representation see \Fig{fig:Heff_misc}(b).

We can now iteratively calculate $E_{[j+1]}=E_{[j]}F_{[j+1]}$ many times until this process has converged. As a
convergence criterion we can use e.g. the ground state energy per site of the renormalized Hamiltonian which is
accumulated in $E^{1}_{[j]}$. Using the converged result as an approximation for $E_{[\ell]}$ we can then easily
determine $H^{\rm eff}_{\rm \triangleleft,\ell+1}$ from \eref{eq:Heff}. 

In the case where MPS matrices $L^{s_{j}}$ and/or MPOs $W_{[j]}^{s_{j}s'_{j}}$ are site dependent for some sites $k\leq
j\leq\ell$, we can first calculate
$E_{[k]}$ up to site $k$ approximately as described above and then calculate the finite product
\begin{equation}
 E_{[\ell]}=E^{\rm init}_{[k]}\prod_{j=k+1}^{\ell}F_{[j]}.
\end{equation} 
Notice that we can in principle even define the left part to be finite altogether, with site dependent matrices
$L^{s_{j\leq\ell}}$ and/or site dependent MPOs $W_{[j]}^{s_{j}s'_{j}}$, such that $E_{[\ell]}=\prod_{j=1}^{\ell}F_{[j]}$
is also a finite product. In this case
one would have to specify left boundary conditions. In our simulations we do not consider this case.

If Method II is used at the right boundary, we use the uniform matrices $R^{s_{j}}$ to construct a renormalized
expression for $\hat{H}_{r,\triangleright}:= \hat{h}_{r,r+1} + \hat{H}_{\rm R}= \sum_{j=r}^{\infty}\hat{h}_{j,j+1}$.

Generally the computational effort for calculating $H^{\rm eff}_{\rm \triangleleft,\ell+1}$ is dictated by the
computational effort for calculating $E_{[\ell]}$. In case of a translation invariant left part, its calculation is
very similar to the \textit{renormalization} steps of an iDMRG simulation \cite{idmrg1,idmrg2} (no eigenvalue/SVD
steps). The number of renormalization steps is dependent on the effective correlation length induced by the uniform MPS
matrices $L^{s_{j}}$. 

In practice it takes about 75 renormalization steps for the TFI model at $h=0.45$ ($m_{0}=30$) and
about 100 steps for the XXZ model at $J_{z}=-2$ ($m_{0}=88$) for convergence in energy up to an accuracy of $10^{-15}$,
where $m_{0}$ is the bond dimension of the ground state MPS representation. The
overall computational effort here is comparable to a few time evolution steps within the CMW.

\subsection{Window Movement}
\label{sec:window_movement}
We describe the window movement by a single site. For a shift by a 2-site unit cell, the same procedure as for a
single site is applied twice. 
 
If no boundary update is used at the left boundary, the matrix $A^{s_{\ell+1}}$ is discarded. If Method II is used,
we incorporate $A^{s_{\ell+1}}$ into the left part by using it to calculate a
renormalized expression for $\hat{H}_{\triangleleft,\ell+2}:= \hat{H}_{\triangleleft,\ell+1} + \hat{h}_{\ell+1,\ell+2}$.
More precisely, we construct $F_{[\ell+1]}$ as defined in \eref{eq:mpo_transfer} using $A^{s_{\ell + 1}}$
\begin{equation}
 F_{[\ell+1]}^{b_{\ell}b_{\ell+1}}:=\sum_{s_{\ell+1}s'_{\ell+1}}W_{[\ell+1],b_{\ell}b_{\ell+1}}^{s_{\ell+1}s'_{\ell+1}}
\bar{A}^ {s_{\ell+1}} \otimes A^{s'_{\ell+1}}.
\end{equation} 
With $E_{[\ell]}$ from earlier calculations we can then construct $E_{[\ell+1]}=E_{[\ell]}F_{[\ell+1]}$, calculate
$H^{\rm eff}_{\rm \triangleleft, \ell+2}$ as defined in \eref{eq:Heff} and determine $U\left(\tau\right)^{\rm
eff}_{\rm \triangleleft, \ell+2}=\exp\left(-\rmi\tau H^{\rm eff}_{\rm \triangleleft, \ell+2}\right)$.

At the right boundary we introduce $R^{s_{r+1}}_{A}$ as a new rightmost matrix $B^{s_{r+1}}$.

After the window has been moved by a single site according to these steps, we redefine $\ell\leftarrow\ell+1$,
$r\leftarrow r+1$ (and we
exchange labels $A\leftrightarrow B$ in the case of iTEBD).

Notice that for the left boundary the dimension of the block basis $\{\ket{a_{\ell}}\}$ can grow with successive window
shifts. An impinging signal can therefore be partly absorbed such that immediate perturbations are considerably
suppressed (see also \ref{sec:reflections}).

We trigger the window shift when the relative change of the bipartite entanglement entropy at some site sufficiently far
away from the right boundary rises above a certain threshold. The margin between this site and the right boundary should
be large in comparison to the correlation length of the initial state such that the exponentially suppressed correlations 
reaching beyond the Lieb-Robinson light cone \cite{lieb-rob} are negligible. For all simulations in the main text we use a
margin of $24$ sites and a threshold of $1\%$.
If known beforehand, the window can also be moved directly with $v_{\rm max}$.

\section{Analytic Results for the TFI Model}
\label{sec:TFI_analytics}
In this appendix, we collect some known results and we derive an exact expression for the transverse magnetization in the TFI model after a Jordan-Wigner excitation.

\subsection{Diagonalization of the Hamiltonian}
The TFI model 
\begin{equation}
 \hat{H}_{TFI}=-\sum_{j}\hat{S}^{x}_{j}\hat{S}^{x}_{j+1} - h\sum_{j}\hat{S}^{z}_{j}
\end{equation}
can be solved exactly \cite{lsm,Pfeuty70} by first transforming to spinless fermionic operators $c^{\dagger}_{j}$,
$c_{j}$ via a Jordan-Wigner (JW) transformation \cite{jw}
\begin{equation}
 \hat{S}^{+}_{j}=\prod_{n<j}\left(1-2c^{\dagger}_{n}c_{n}\right)c^{\dagger}_{j},\qquad
 \hat{S}^{-}_{j}=\prod_{n<j}\left(1-2c^{\dagger}_{n}c_{n}\right)c_{j},
 \label{eq:JW}
\end{equation} 
where $\hat{S}^{+}_{j}$ and $\hat{S}^{-}_{j}$ are the spin raising and lowering operators. With
$\S^{x}_{j}=\frac{1}{2}\left(\S^{+}_{j}+\S^{-}_{j}\right)$ and $\S^{z}_{j}=\S^{+}_{j}\S^{-}_{j}-
\frac{1}{2}$ the Hamiltonian becomes
\begin{equation}
\hat{H}_{TFI}=-\frac{1}{4}\sum_{j}\left(c^{\dagger}_{j} - c_{j}\right)\left(c^{\dagger}_{j+1} + c_{j+1}\right)- h\sum_{j}\left(c^{\dagger}_{j}c_{j}-\frac{1}{2}\right).
\end{equation}
Here we have already taken the thermodynamic limit while considering periodic
boundary conditions (A boundary term arising from the JW transformation and periodic boundary conditions is
neglected as it is of the order $\Or(1/L)$ where $L$ is the system size).

A subsequent Bogoliubov transformation \cite{bog} to fermionic operators $\eta_{k}$, $\eta^{\dagger}_{k}$ in momentum
space
\begin{equation}
 c_{j}=\frac{1}{\sqrt{2\pi}}\int_{-\pi}^{\pi} \rmd k\, \rme^{\rmi kj}\left(a_{k}\eta_{k} - \rmi b_{k}\eta_{-k}^{\dagger}\right)
 \label{eq:bogoliubov}
\end{equation}
then diagonalizes the Hamiltonian. 
The coefficients $a_{k}$ and $b_{k}$ are real and satisfy
\begin{equation}
 a_{-k}=a_{k}, \qquad b_{-k}=-b_{k}, \qquad a_{k}^2 + b_{k}^2 = 1
 \end{equation}
and can be determined as
\begin{eqnarray}
 a_{k} &=& \frac{\varepsilon_{k} - \frac{1}{2}\cos(k) - h}{\sqrt{2\varepsilon_{k}\left(\varepsilon_{k} -
\frac{1}{2}\cos(k) - h\right)}}\\
b_{k}&=&-\frac{\frac{1}{2}\sin(k)}{\sqrt{2\varepsilon_{k}\left(\varepsilon_{k} -
\frac{1}{2}\cos(k) - h\right)}}\\
\varepsilon_{k}&=&\sqrt{\frac{1}{4}+h\cos(k)+h^2}.
\end{eqnarray}
The Hamiltonian then reads
\begin{equation}
 \hat{H}_{TFI}=\frac{1}{2\pi}\int_{-\pi}^{\pi}\rmd k\,\varepsilon_{k}\left(\eta^{\dagger}_{k}\eta_{k}-\frac{1}{2}\right)
\end{equation} 
and the ground state corresponds to the vacuum state $\ket{0}$ in terms of the fermionic operators $\eta_{k}$ and
$\eta^{\dagger}_{k}$.
\subsection{Signal Velocity in the TFI model}
\label{sec:TFI_velocity}
The propagation of a signal induced on top of the ground state $\ket{0}$ of the TFI model
can be understood as the excitation and propagation of a superposition of non-interacting particles
with momenta $k$ and corresponding energies $\varepsilon_{k}$ created by  $\eta^{\dagger}_{k}$. In this picture, the maximum velocity $v_{\rm max}$ of
the signal can be exactly calculated as the maximum of the group velocity
\begin{equation}
v_{k}:=\frac{\rmd\varepsilon_{k}}{\rmd k}=\frac{h}{2}\frac{\sin(k)}{\varepsilon_{k}}.
\end{equation}
A short calculation shows that $v_{k}$ takes its extrema at $\cos(k)=2h$ and $\cos(k)=\frac{1}{2h}$,
which gives
\begin{equation}
 v_{\rm max}=
 \left\{ \begin{array}{lll}
  h&, &h\leq h_{\rm crit}\\
  h_{\rm crit}&, &h\geq h_{\rm crit}
 \end{array} \right.
\end{equation} 
where $h_{\rm crit}=\frac{1}{2}$.

\subsection{Analytic Results for a JW Excitation}
\label{sec:JW_anal_res}
Consider a Jordan-Wigner (JW) excitation at site $\ell$ on top of the thermodynamic limit ground state $\ket{0}$, defined as
\begin{eqnarray}
\ket{\psi}_{\ell}&=&\left(c^{\dagger}_{\ell} + c_{\ell}\right)\ket{0}\\
&=&\frac{1}{\sqrt{2\pi}}\int_{-\pi}^{\pi} \rmd k\,\rme^{-\rmi k\ell}\left(a_{k}+\rmi b_{k}\right)\eta^{\dagger}_{k}\ket{0}.
\label{eq:JW_psi0}
\end{eqnarray} 

Using the results from the previous sections for the TFI model, the time evolution of the magnetization in \textit{z}
after such an excitation
\begin{equation}
 \braket{\hat{S}^{z}(n,t)}_{\ell} = \bra{\psi}_{\ell}c^{\dagger}_{n}(t)c_{n}(t)\ket{\psi}_{\ell} - \frac{1}{2}.
 \label{eq:JW_magn_timeevo1}
\end{equation}
can be calculated analytically.

Solving the Heisenberg equation of motion for $\eta_{k}(t)$ yields $\eta_{k}(t)=\rme^{-\rmi\varepsilon_k t}\eta_{k}$. 
Using \eref{eq:bogoliubov} then allows to write

\begin{equation}
 c_{n}(t)=\frac{1}{\sqrt{2\pi}}\int_{-\pi}^{\pi}\rmd k\,\rme^{\rmi kn}\left(\alpha_{k}(t)\,\eta_{k} +
\beta_{k}(t)\,\eta_{-k}^{\dagger}\right)
\label{eq:JW_ops_timeevo}
\end{equation} 
with $\alpha_{k}(t)=a_{k}\rme^{-\rmi\varepsilon_k t}$ and $\beta_{k}(t)=-\rmi b_{k}\rme^{\rmi\varepsilon_k t}$. Plugging
\eref{eq:JW_ops_timeevo} and \eref{eq:JW_psi0} into \eref{eq:JW_magn_timeevo1} yields after some calculation

\begin{equation}
 \boxed{\braket{\hat{S}^{z}(n,t)}_{\ell} = S^{z}_{\rm GS} + 
\frac{I^{A}_{\ell}(n,t) - I^{B}_{\ell}(n,t)}{(2\pi)^2},}
  \label{eq:JW_magn_timeevo2}
\end{equation} 
where
\begin{equation}
 S^{z}_{\rm GS} = \bra{0}\S^{z}_{j}\ket{0} = \frac{1}{2\pi}\int_{-\pi}^{\pi}\rmd k\,\left|b_{k}\right|^2 - \frac{1}{2}
\end{equation} 
is the ground state magnetization and
\begin{eqnarray}
 I^{A}_{\ell}(n,t)&=\Big|\int_{-\pi}^{\pi}\rmd k\,\rme^{-\rmi\left[ k(\ell-n) + \varepsilon_k t\right]} 
 \left(a_{k} + \rmi b_{k}\right)a_{k} \Big|^2\\
 I^{B}_{\ell}(n,t)&=\Big| \int_{-\pi}^{\pi}\rmd k\,\rme^{-\rmi\left[ k(\ell-n) - \varepsilon_k t\right]} \left(a_{k} +
\rmi b_{k}\right)b_{k} \Big|^2. 
\end{eqnarray}

In \ref{sec:precision} we use \eref{eq:JW_magn_timeevo2} to compare with results obtained from a CMW simulation. 

\section{Analytic Results for the XXZ Model}
We derive an exact expression for the group velocities in the XXZ model and calculate the signal velocity $v_{\rm max}$.

\label{sec:XXZ_analytics}
\subsection{Bethe Ansatz Solution for the ground state}
The XXZ model defined by the Hamiltonian
\begin{equation}
 H_{\rm XXZ} = -\sum_{j}\S^{x}_{j}\S^{x}_{j+1} + \S^{y}_{j}\S^{y}_{j+1} + \Delta\S^{z}_{j}\S^{z}_{j+1}
\end{equation} 
can be solved e.g. by means of the coordinate Bethe ansatz \cite{Bethe}. 

We seek solutions for the ground state and elementary excitations of the XXZ antiferromagnet with $\Delta<-1$ in the
thermodynamic limit, which can be found e.g. in \cite{BetheXXZ_CG}.

In the thermodynamic limit the roots of the Bethe equations become dense and their distribution for the ground state is
characterized by a density function $g_{0}(x)$, which for $\Delta<-1$ satisfies the following integral equation

\begin{eqnarray}
 g_{0}(x)&+&\frac{\sinh(2\Phi)}{2\pi}\int_{-\pi}^{\pi}\frac{g_{0}(x')\rmd x'}{\cosh(2\Phi) - \cos(x - x')}\\
 ~&=&\frac{2\sinh(\Phi)}{\cosh(\Phi) - \cos(x)},
\end{eqnarray}
where $\cosh(\Phi) = -\Delta$.

The solution to this integral equation is given by
\begin{equation}
 g_{0}(x) = \frac{2K(m_{0})}{\pi}\dn\left(\frac{K(m_{0})}{\pi}x,m_{0}\right),\quad -\pi<x<\pi.
\end{equation} 
Here $\dn(x,m)$ is a Jacobian Elliptic Function \cite{MathFun}, $K(m)$ the Complete Elliptic Integral
of the first kind 
\begin{equation}
 K(m) = \int_{0}^{\frac{\pi}{2}}\frac{\rmd x}{\sqrt{1-m\sin^2 x}}
\end{equation} 
and the parameter $m_{0}$ the solution of the equation
\begin{equation}
 \frac{K(m_{0})}{K(1-m_{0})} = \frac{\pi}{\Phi}.
\end{equation} 

The root density $g_{0}(x)$ can then be used to calculate various quantities such as the ground state energy and
elementary excitations.
\subsection{Signal Velocity in the XXZ model}
To calculate the maximum signal velocity $v_{\rm max}$ as a function of $\Delta$ we first determine the dispersion
relation $\varepsilon_{k}$ for the elementary excitations. As for the TFI model in \ref{sec:TFI_velocity} we then
obtain $v_{\rm max}$ as the maximum of the group velocity $v_{k}=\frac{d\varepsilon_{k}}{dk}$. 

The dispersion of the elementary excitations is given by \cite{BetheXXZ_CG}
\begin{equation}
 \varepsilon_k = \frac{1}{2}\sinh(\Phi)\left[g_{0}(x_{0}(k)) - g_{0}(\pi)\right] + G(\Delta),
 \label{eq:XXZ_e}
\end{equation} 
where $G(\Delta)$ is the finite energy gap present in this phase and $x_{0}(k)$ has to be determined by inverting 
\begin{equation}
 k(x_{0}) = \frac{1}{2}\left(\pi - \int_{0}^{x_{0}}g_{0}(x)\rmd x\right),\quad -\pi\leq x_{0} \leq \pi
 \label{eq:XXZ_k}
\end{equation}
for a given momentum $k$.

\begin{table}[t]
\centering
 \begin{tabular}{|c|c||c|c|}
 \hline
  $\Delta$ & $v_{\rm max}$ & $\Delta$ & $v_{\rm max}$\\
  \hline
  -1.5 &  1.78173404 & -3.5 &  1.95920177\\
  -2.0 & 1.87559502 & -4.0 & 1.96875800\\
  -2.5 & 1.92014492 & -4.5 & 1.97531255\\
  -3.0 & 1.94449113 & -5 & 1.98000206\\
  \hline
 \end{tabular} 
\caption{Values for the maximum signal velocity $v_{\rm max}$ in the XXZ model for various values of the interaction
strength $\Delta<-1$. These values are obtained from numerically finding the maximum of \eref{eq:XXZ_vmax} with
a numerical precision of $10^{-8}$.
}
\label{tab:XXZ_vmax}
\end{table} 

From this we can calculate the group velocity
\begin{equation}
 v_{k} = \frac{\rmd\varepsilon_{k}}{\rmd k} = \frac{\rmd\varepsilon_{k}}{\rmd x_{0}}\left(\frac{\rmd k}{\rmd x_{0}}\right)^{-1}
\end{equation} 
where we need 
\begin{equation}
\frac{\rmd\varepsilon_{k}}{\rmd x_{0}} = \frac{1}{2}\sinh(\Phi) \frac{\rmd g_{0}}{\rmd x_{0}},\qquad
\frac{\rmd k}{\rmd x_{0}} = -\frac{g_{0}(x_{0})}{2}.
\end{equation}
Using some properties of Jacobian Elliptic Functions \cite{MathFun} and defining $K_{0}=K(m_{0})$ we get
\begin{equation}
 \frac{\rmd g_{0}}{\rmd x_{0}} = -2\, m_{0} \left(\frac{K_{0}}{\pi}\right)^{2}
\sn\left(\frac{K_{0}x_{0}}{\pi},m_{0}\right)\cn\left(\frac{K_{0}x_{0}}{\pi},m_{0}\right)
\end{equation} 
where $\sn(x,m)$ and $\cn(x,m)$ are also Jacobian Elliptic functions. Defining
\begin{equation}
 S(x_{0},K_{0},m_{0}):=\sn\left(\frac{K_{0}x_{0}}{\pi},m_{0}\right)\cn\left(\frac{K_{0}x_{0}}{\pi},m_{0}\right)
\end{equation} 
we can then write
\begin{equation}
\boxed{
 v_{k} = \frac{2\,m_{0}\sinh(\Phi)}{g_{0}(x_{0})}\left(\frac{K_{0}}{\pi}\right)^{2}S(x_{0},K_{0},m_{0}).
 }
 \label{eq:XXZ_vmax}
\end{equation} 

We determine the maximum of this function numerically to get $v_{\rm max}$ as a function of $\Delta$. The values
of $v_{\rm max}$ for various interaction strengths $\Delta<-1$ are listed in table
\ref{tab:XXZ_vmax} and are obtained with a numerical precision of $10^{-8}$.

\section{Test of precision of results}
\label{sec:precision}
To assess the accuracy of the CMW approach, we compare it with a
reference system on a very large lattice and with exact results obtained in \ref{sec:JW_anal_res} . 
We investigate simulations of a JW excitation on top of the infinite system ground state in
the TFI model at $h=0.45$ for windows of different sizes, different boundary update methods and different margins between
signal and right boundary for triggering the window shift. Note that the 
correlation length 
of this system is $\xi\approx4.36$ sites. It can be obtained from the second
largest eigenvalue in magnitude $\lambda_{2}$ of the MPS transfer matrix 
$T=\sum_{s}\bar{A}^{s} \otimes A^{s}$
as $\xi^{-1}=-\log(\left|\lambda_{2}\right|)$
\cite{MPS1,OstlundRommer}.
For all simulations we use second order Suzuki-Trotter decomposition with time step
$\tau=0.002$ and maximum bond dimension $m_{\rm max}=120$. 
These are the same simulation parameters
as used for the investigation of a JW excitation in the TFI model in the main text.

The reference simulation is also performed using the CMW algorithm, but starting with the translation invariant initial state inside a non-moving
window of very large size of $N=1000$ sites. This means that the window is never shifted. Boundary effects are
removed by using Method I for \textit{both} boundaries. For the reference simulation we perform time evolution up to
$t=800$, such that the signal induced in the centre of the system at $t=0$ does not reach the boundaries. For a
plot of the reference simulation see \Fig{fig:refsim}. 
There we show the transverse magnetization $S^{z}(n,t)$ and the
bipartite entanglement entropy $S_{\rm ent}(n,t)$.
It can be seen that boundary effects are indeed removed for
the non-moving window with method I 
(otherwise disturbances would constantly radiate from the boundaries) 
and that the signal is still about $\approx 150$ sites away from the boundaries at $t=800$.

\begin{figure}[t]
 \centering
 \includegraphics[width=\linewidth,keepaspectratio=true]{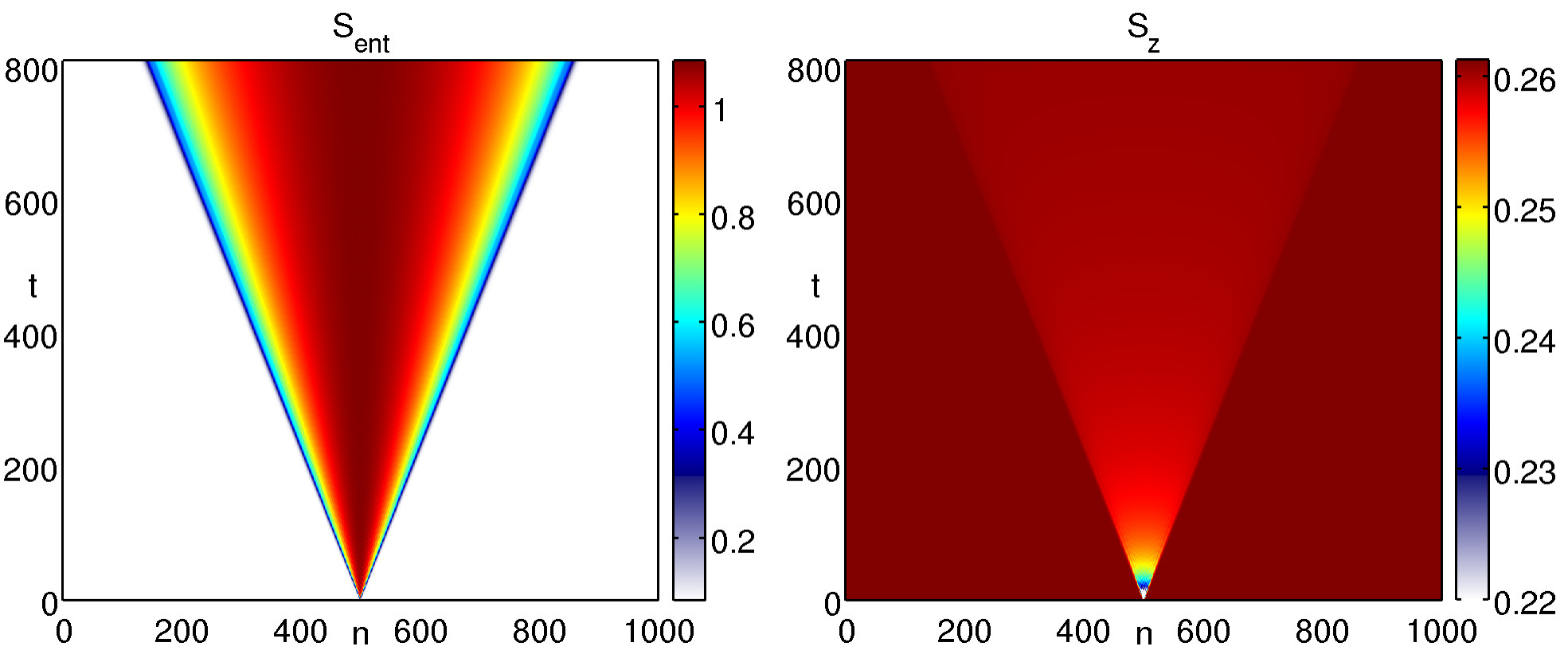}
 \caption{Plot of the reference simulation used for
comparison to CMW simulations with different setups described in table \ref{tab:setups}. The bipartite entanglement
entropy $S_{\rm ent}(n,t)$ and the transverse magnetization $S^{z}(n,t)$ of a JW excitation in the TFI model at $h=0.45$ are shown. 
We use a non-moving window (boundary effects are completely removed by using Method I
at both boundaries) with $N=1000$ sites and perform time evolution using second order Suzuki-Trotter decomposition with
time step $\tau=0.002$ and maximum bond dimension $m_{\rm max}=120$ up to $t=800$.
}
 \label{fig:refsim}
\end{figure}

\begin{figure}[p]
 \centering
 \includegraphics[width=0.9\linewidth,keepaspectratio=true]{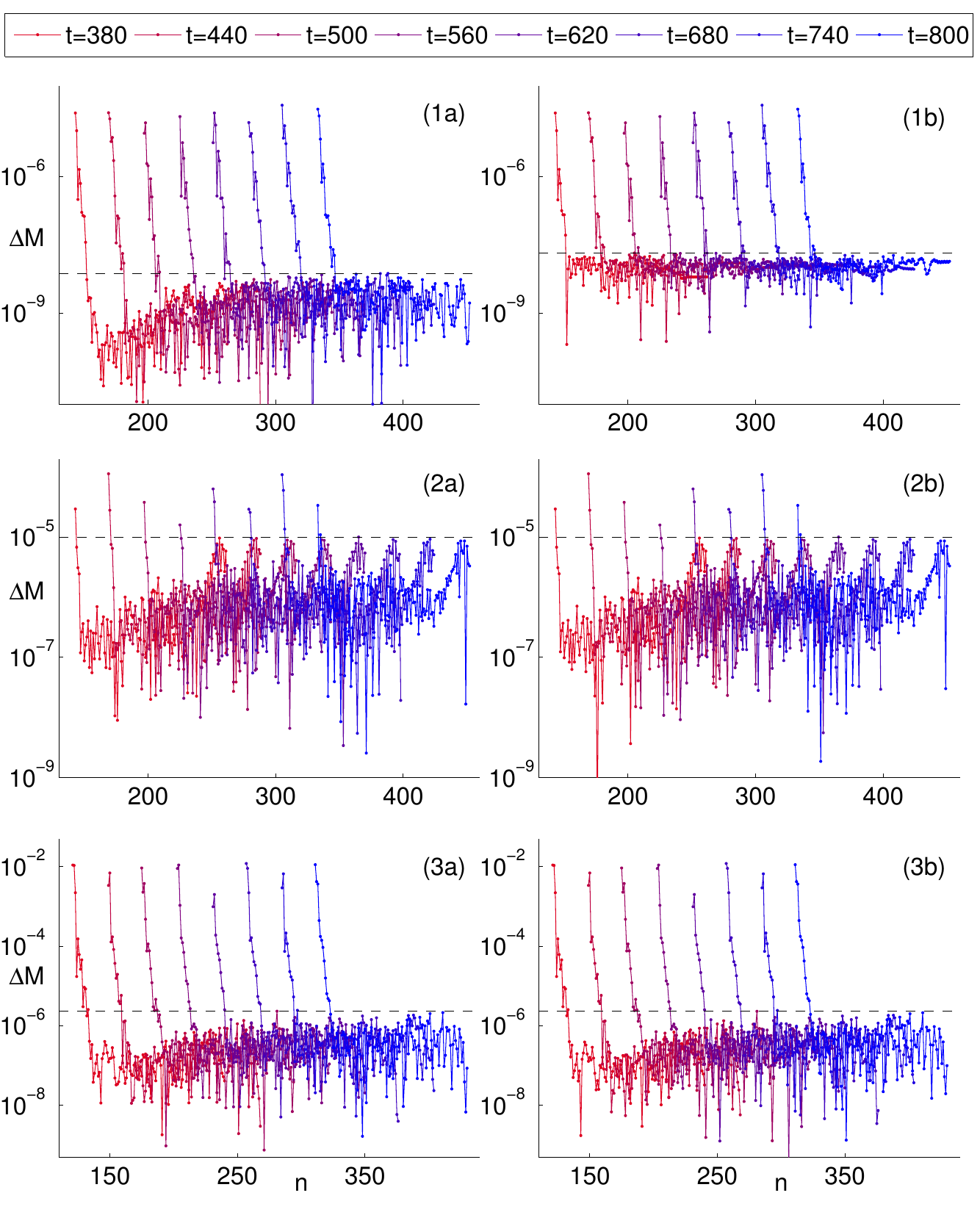}
 \caption{Comparison of results from different selected setups described in table
\ref{tab:setups} to (a) results from a \textit{reference simulation} (left column) and (b) \textit{analytic results}
\eref{eq:JW_magn_timeevo2} (right column). We plot the absolute differences in measured transverse magnetization $\Delta
M(n,t)$ \eref{eq:Mdiff} vs. absolute position $n$ for various times $380\leq t\leq 800$ . The black dashed lines
mark the values of the largest absolute differences inside the CMW away from the left boundary.
We note that comparisons of the \textit{longitudinal} magnetization and the entanglement entropy to the reference simulation yield very similar results (not shown).
}
 \label{fig:cmw_compare_setups}
\end{figure}

We compare results from setups with different CMW sizes $N$ and different numbers of margin sites between signal and right boundary (in sites,
see \ref{sec:window_movement}), as well as different setups for using Method I and II for the updates at the left
boundary (LB) and right boundary (RB). 
We find that the accuracy of the simulation strongly depends on the boundary update method used at the right
boundary and the margin between signal and right boundary, whereas the window size $N$ has virtually no impact on the
accuracy. For a selection of compared setups see table \ref{tab:setups}.

For comparison we will consider the transverse magnetization, since only this quantity is available analytically. 
We display the absolute value of the 
difference in transverse magnetization, 
\begin{equation}
\Delta M(n,t) := \big|S^{z}(n,t)_{\rm anal./ref.} -
S^{z}(n,t)_{[j]}\big|
\label{eq:Mdiff}
\end{equation}
between the reference simulation (ref.) or the analytic result (anal.) \eref{eq:JW_magn_timeevo2}, respectively,  and the individual setups
$[j]$.
We plot this quantity vs.
absolute position $n$ at various times $380\leq t\leq 800$  in \Fig{fig:cmw_compare_setups}.

For other observables, analytic results are not available, but we can compare to the reference simulation.
We find that comparison of the
magnetization in $x$ direction $S^{x}(n,t)$ and of the bipartite entanglement
entropy $S_{\rm ent}(n,t)$ to the reference simulation yield results that look very similar to
\Fig{fig:cmw_compare_setups} and the obtained absolute differences are also of the same orders of magnitude.
We note in addition that comparison  between left and right column in \Fig{fig:cmw_compare_setups} 
confirms the absence of boundary effects in the reference simulation to high precision.

In the following we discuss the comparisons of the 3 cases listed in table \ref{tab:setups}. 

Case (1) corresponds to the same CMW setup as used for data analysis in the main text. Comparison  to
the reference simulation yields differences of at most $\Or(10^{-9})$, whereas a comparison to  analytic
results yields differences of at most $\Or(10^{-8})$ everywhere inside the CMW. Around the left
boundary differences are of $\Or(10^{-5})$ for both cases due to perturbations arising from the impinging left
going part of the signal. These perturbations however remain confined around the left boundary at all times.

In case (2), where Method II is used at the \textit{right} boundary, differences inside the CMW rise up to
$\Or(10^{-5})$ for both comparisons, i.e. they are
considerably higher by about 3-4 orders of magnitude in comparison to case (1), where Method I is used. This can be
explained by the fact that Method II breaks the structure of the Suzuki-Trotter decomposition at the boundary, which
introduces additional perturbations. These perturbations can in principle be reduced by using higher order
Suzuki-Trotter decompositions and smaller time steps and thus increasing computational effort, but they are always
present. Method I however is completely devoid of this kind of perturbations. 
Also, the renormalized Hamiltonian $H^{\rm eff}_{r \triangleright}$ necessary for Method II is only
calculated up to a finite precision. We however find the perturbations to be largely
independent of the precision used to calculate $H^{\rm eff}_{r \triangleright}$ as described in \ref{sec:ren_ham}.
We conclude that using Method I at the right boundary yields results
which are better by about 3-4 orders of magnitude in precision than using Method II when employing second order
Suzuki-Trotter decomposition with a time step of $\tau=0.002$. 

In case (3) the left part has been disconnected from the CMW altogether by setting
$\hat{h}_{\ell,\ell+1}=0$ (``cut'') as described in the main text.
Also the margin between signal and right boundary is reduced to 3 sites. Due to the cut, perturbations
around the left boundary are now considerably higher and go up to $\Or(10^{-2})$ both for the comparison to
analytic results and the reference simulation. These perturbations however again remain confined around the left
boundary at all times. Differences inside the CMW are now $\Or(10^{-6})$ for both comparisons. This can be
explained by the fact that the margin of 3 sites is now smaller than the correlation length $\xi\approx4.36$ and the
exponentially suppressed correlations reaching beyond the Lieb-Robinson light cone \cite{lieb-rob} induce perturbations at the right
boundary.

\begin{table}[t]
 \centering
\begin{tabular}{|c|cccccc|}
\hline
 & $N$ & margin& LB & RB & $P_{\rm ref.}$ & $P_{\rm anal.}$\\
 \hline
 (1) & 120 & 24 & II & I & $7.3\times10^{-9}$ & $2.1\times10^{-8}$ \\
 (2) & 120 & 24 & II & II & $1.0\times10^{-5}$ & $1.0\times10^{-5}$ \\
 \hline
 (3) & 120 & 3 & cut & I & $2.3\times10^{-6}$ & $2.3\times10^{-6}$ \\
 \hline
 ref. & 1000 & - & I & I& - & $2.0\times10^{-8}$\\
 \hline
\end{tabular} 
\caption{Precision of different CMW simulation setups, for the case of a JW excitation on top of the infinite system
ground state in the TFI model at $h=0.45$. We compare CMW results on $N=120$ sites with analytic results (anal.)
and with results from a reference simulation on $N=1000$ sites (ref.). 
"Margin" specifies the number of sites kept between the signal and the right boundary of the CMW as explained in \ref{sec:window_movement}. 
The precision $P_{\rm ref./anal.}$ is the resulting maximum absolute difference in transverse magnetizations \eref{eq:Mdiff} inside the CMW away from the left boundary,
between the CMW simulation and the reference simulation or analytic result (black dashed lines in \Fig{fig:cmw_compare_setups}).
All simulations were performed using $m_{\rm max}=120$ and second order Suzuki-Trotter decomposition with time
step $\tau=0.002$ up to $t=800$. Case (1) corresponds to the setup used for data analysis in the main text. For (3),
``cut'' means that the CMW is disconnected from the left part by setting $\hat{h}_{\ell,\ell+1}=0$, corresponding to
the simplest to implement setup, as described in the main text. A comparison between cases (1) and (2) shows that
Method I yields very precise results, better by several orders of magnitude than Method II. }
\label{tab:setups}
\end{table}

In conclusion, both Method I (Uniform Update) and Method II (Renormalized Update) work quite well.
Furthermore, the easy to implement Method I yields results with a precision of about $10^{-8}$, 
still better by several orders of magnitude than Method II when used at the right boundary. 
For the methods to work, 
the margin between signal and right boundary needs to be considerably larger than the correlation length.
%
At the left boundary the easiest approach, a simple cut, already works well when the very rear of the CMW is not of interest. 


Overall we have shown that the error produced by the CMW approach, especially with Method I, is very  small and remains virtually constant
for large times during the simulation when the margin between signal and right boundary is kept sufficiently larger than
the correlation length in the initial state.

\section{Unscaled Time Evolution Results}
\label{sec:unscaled_results}
In this Section we show time evolution results before scaling for the TFI model and the XXZ model, for the signals
investigated in the main text.
\subsection{TFI model}

\begin{figure}[t]
 \centering
 \includegraphics[width=0.9\linewidth,keepaspectratio=true]{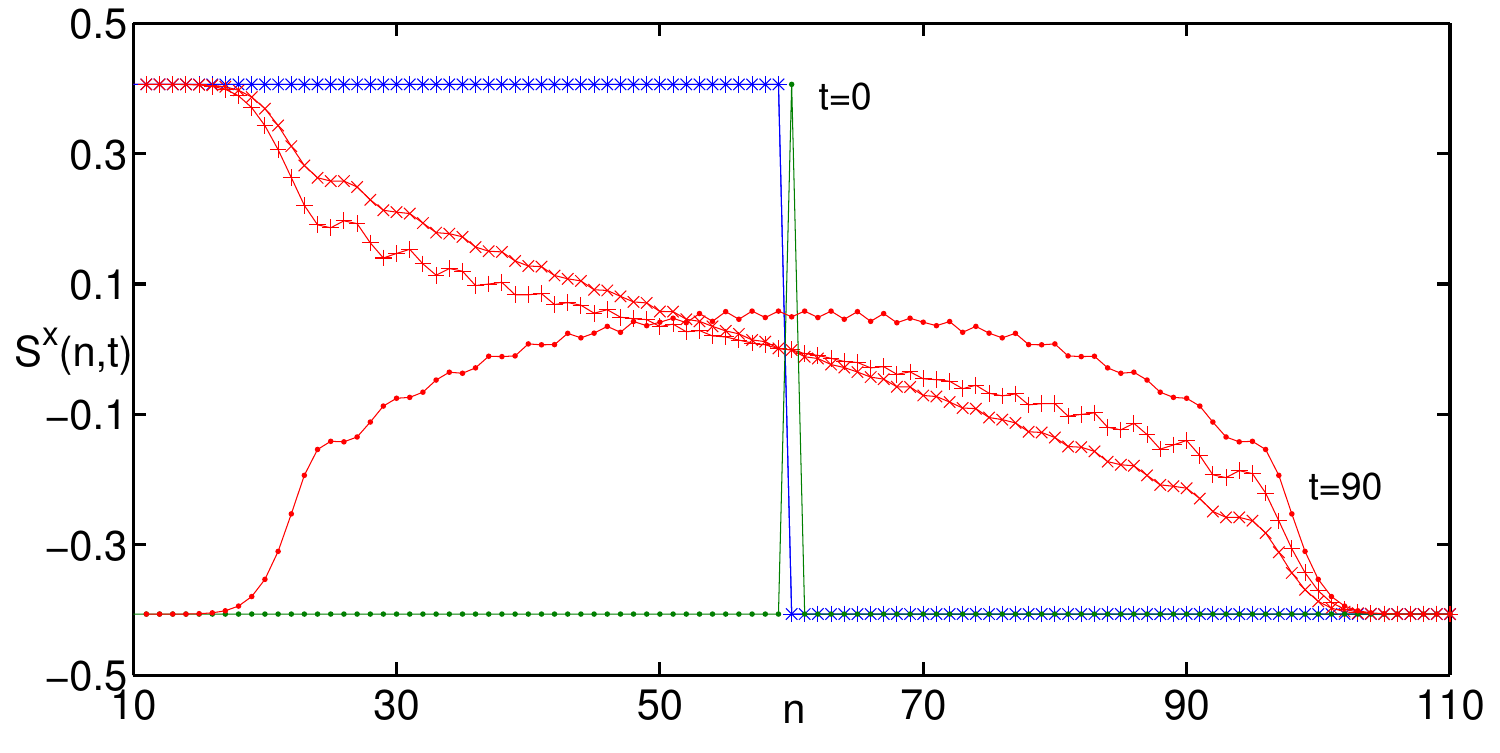}
 \caption{Unscaled magnetization $S^{x}(n,t)$ of the TFI model at $h=0.45$ vs. absolute position $n$.
At $t=90$ (red) from top to bottom on the right side: single spin flip in $x$-direction (dot symbols),
a domain wall excitation (+ symbols), and a JW excitation (x symbols).
The initial state at $t=0$ was a delta spike (green) for the single spin flip and a step function (blue)
for the two other excitations.
 }
 \label{fig:TFI_overview}
\end{figure}

\begin{figure}[t]
 \centering
 \includegraphics[width=0.9\linewidth,keepaspectratio=true]{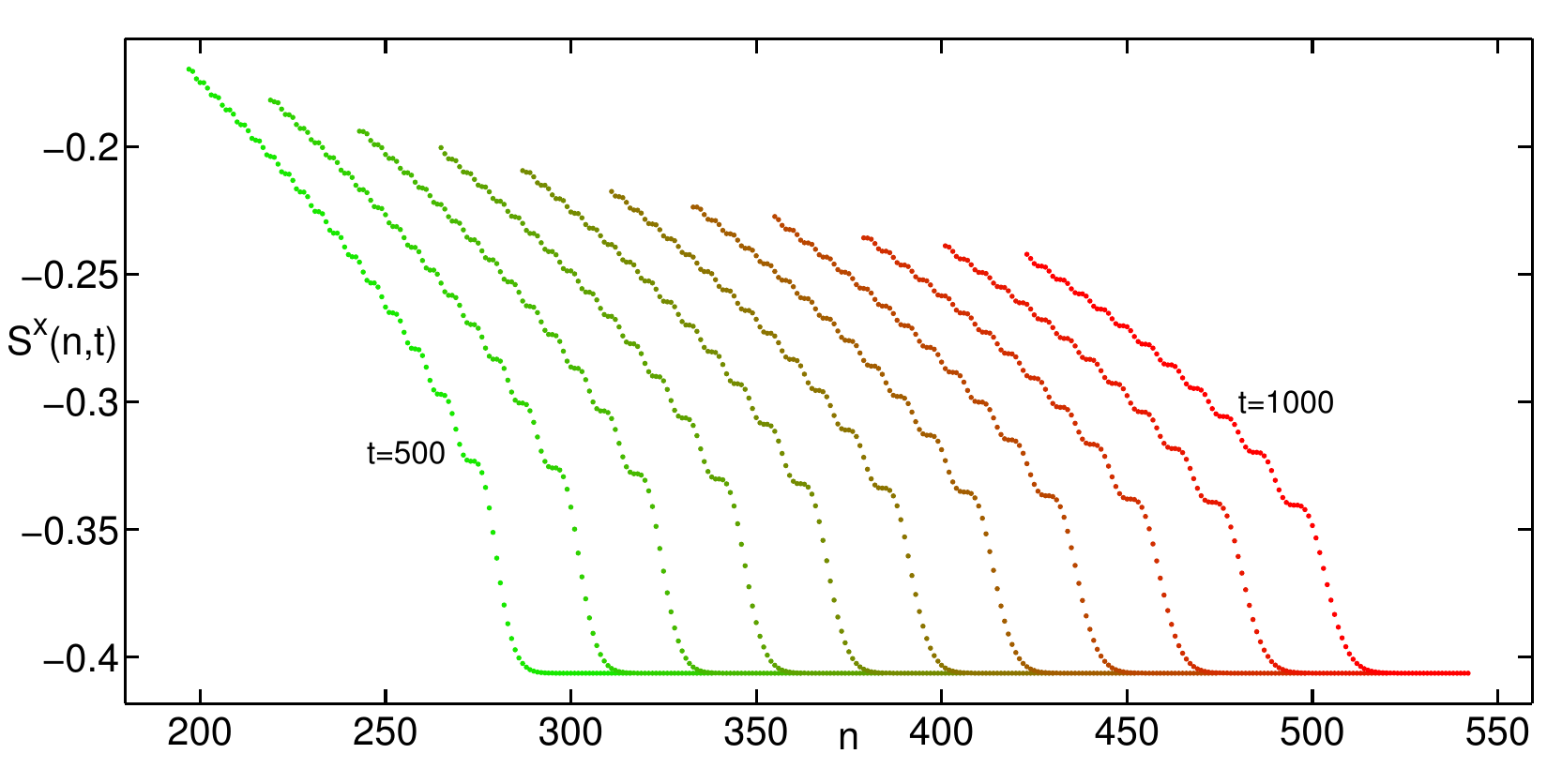}
 \caption{Unscaled magnetization $S^{x}(n,t)$ of the TFI model at $h=0.45$ vs. absolute position $n$ 
after a JW excitation for times $500<t<1000$ in time steps of
$50$.}
 \label{fig:TFI_slice}
\end{figure}

\begin{figure}[t]
 \centering
 \includegraphics[width=0.85\linewidth,keepaspectratio=true]{\appendixfigpath/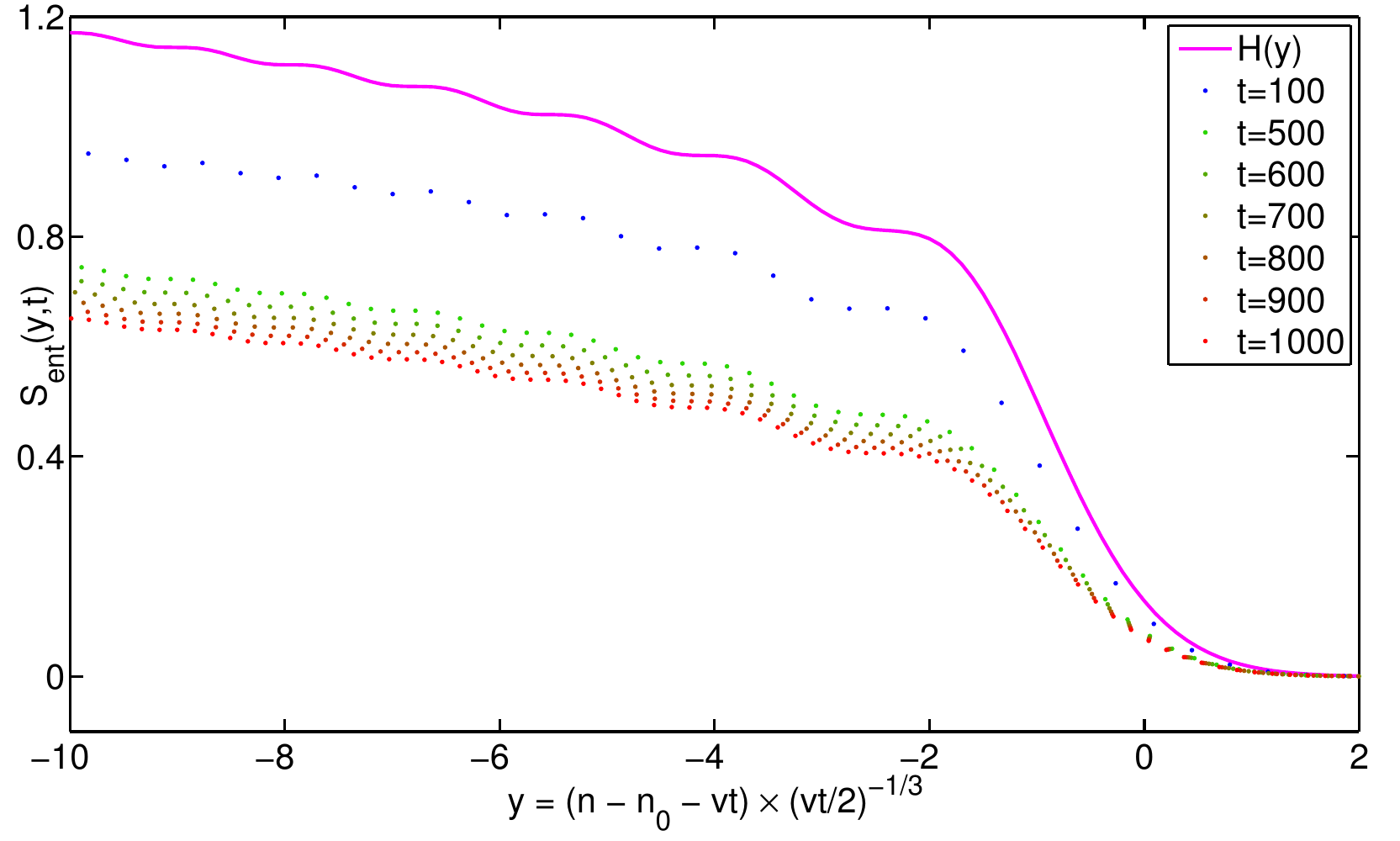}
 \caption{Unscaled bipartite entanglement entropy $S_{\rm ent}^{\rm TFI}(y,t)$ of the TFI model $h=0.45$ after a JW excitation vs. \textit{scaled} position $y$, which is at all times smaller than the entanglement entropy $S_{\rm ent}^{\rm TB}(y,t)$
 for tight binding fermions, which approaches the scaling function $H(y)$ \textit{without} any scaling (c.f. figure 3 in \cite{Eisler13})
 }
 \label{fig:TFI_Sent_unscaled}
\end{figure}

In our simulations we use a Trotter step size of $\tau=0.002$ and a maximum matrix dimension $m_{\rm max}=120$.
The unscaled magnetization $S^{x}(n,t)$ in the TFI model 
for the three different quenches employed is shown in \Fig{fig:TFI_overview} for times $t=0$ and $t=90$.
The global shapes are quite different,
while developing plateaus are visible for all three quench types at $t=90$.
It can also be seen that around the 
signal front, the magnetization of a single spin flip is always larger than of a domain wall, which in turn is 
always greater than the magnetization of a JW excitation. This fact is reflected in the different values for the 
constant $C$ in \Fig{fig:TFI_combined} of the main text.

The unscaled $S^{x}(n,t)$ at $h=0.45$ after a JW excitation in the
infinite system ground state vs. absolute position $n$ at large times
$500<t<1000$ is shown in \Fig{fig:TFI_slice}. The ballistic propagation of the signal front as well as  
magnetization steps near the front are clearly
visible. 
No such steps appear in the transverse magnetization.
A scaling behaviour of the magnitude and distance to the signal front of the steps can be conjectured. 
This scaling behaviour is discussed in detail in the main text.

Other observables and signals, such as single spin flip and domain wall excitations qualitatively show the same 
propagation, shape and step structure. Their scaling behaviour however varies in scaling exponents and quality
with varying field strength $h$.

We also show the bipartite entanglement entropy $S^{\rm TFI}_{\rm ent}(n,t)$ after a JW excitation at $h=0.45$ in \Fig{fig:TFI_Sent_unscaled}. 
It is smaller than the entanglement entropy for tight binding fermions after a domain wall excitation at all times. 
In the latter case, $S^{\rm TB}_{\rm ent}(n,t)$ approaches the asymptotic 
scaling function $H(y)$ \textit{without} any scaling in time. In fact, $S^{\rm TFI}_{\rm ent}(n,t)$ decreases in time, whereas $S^{\rm TB}_{\rm ent}(n,t)$ approaches $H(y)$ from below. 
The exact relation between the scaled excess longitudinal magnetization $M(n,t)=(S^{x}(n,t)-S^{x}_{\rm GS})/{|2S^{x}_{\rm GS}|}$ and the fermion density $N_{TB}(n,t)$ described in the main text therefore does 
not carry over to the entanglement entropy.

\subsection{XXZ model}

\begin{figure}[t]
 \centering
 \includegraphics[width=0.9\linewidth,keepaspectratio=true]{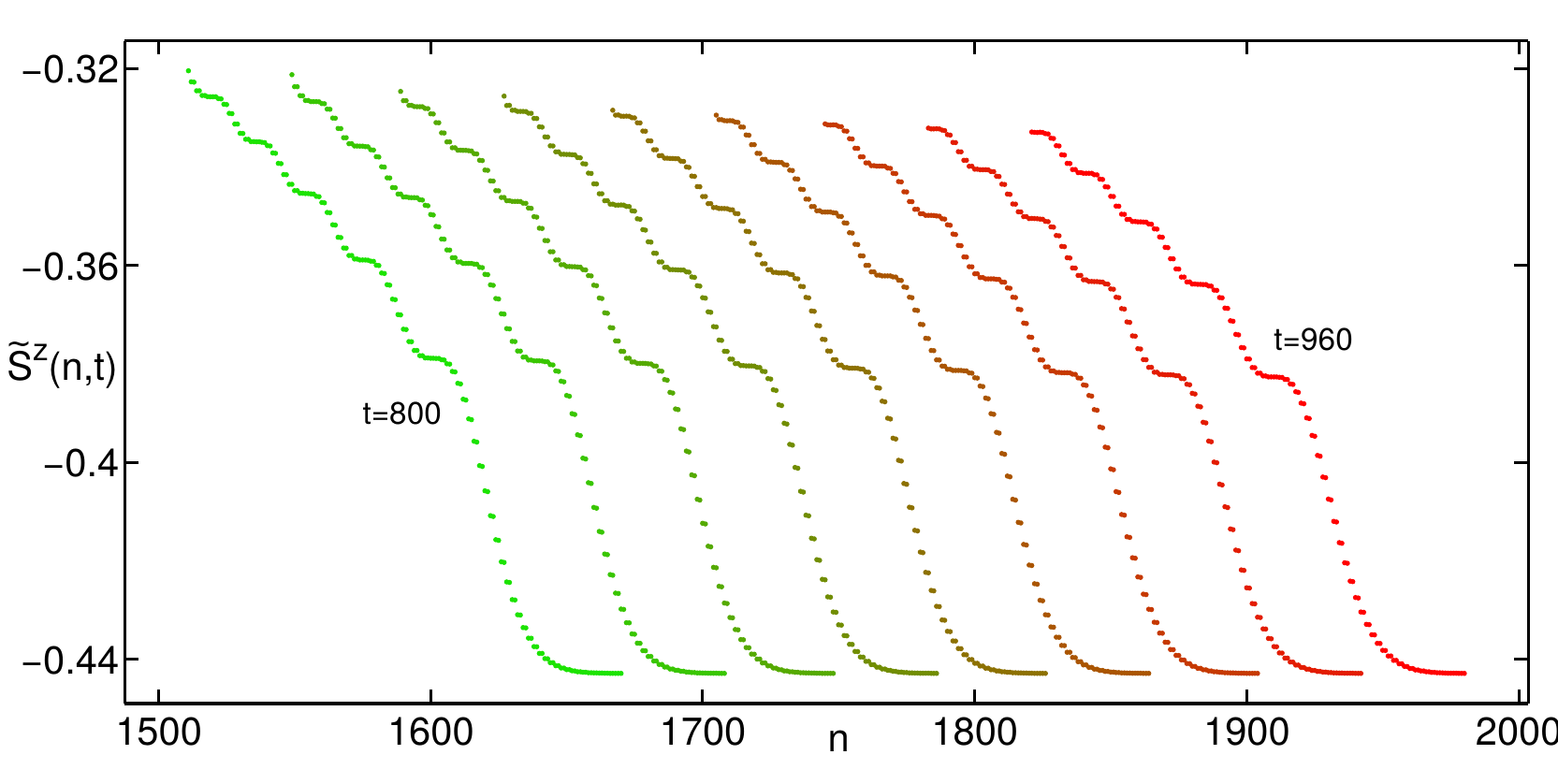}
 \caption{Unscaled staggered magnetization $\tilde{S}^{z}(n,t)$ of the XXZ model at $\Delta=-3$ after a JW
excitation vs. absolute position $n$ at times
$800<t<960$ in time steps of $20$.}
 \label{fig:XXZ_slice}
\end{figure}

For the XXZ simulations we use a Trotter step size of $\tau=0.01$ and maximum matrix dimensions of
$m_{\rm max}=150$ for $\Delta=-4$, $m_{\rm max}=160$ for $\Delta=-3$ and $m_{\rm max}=180$ for $\Delta=-2$.
We show a representative plot of the unscaled staggered magnetization $\tilde{S}^{z}(n,t)=(-1)^{n}S^{z}(n,t)$
at $\Delta=-3$ after a JW excitation on top of the infinite system ground state vs. absolute position $n$ at various times
$800<t<960$ in \Fig{fig:XXZ_slice}.
Again we observe a ballistic propagation of the signal front as well as
magnetization steps near the front as in the TFI model case. For the XXZ model an additional micro step
structure appears due to ``pairing'' of neighbouring sites, which is due to the spinon nature of the elementary excitations
created by the signal (see \Sec{sec:XXZ_results}).

The scaling behaviour of the larger step structure is investigated in detail in the main text. The overall shape of the
unscaled staggered magnetization $\tilde{S}^{z}(n,t)$ looks similar to the shape of the longitudinal magnetization $S^{x}(n,t)$
of the TFI model with a JW excitation as shown in \Fig{fig:TFI_overview}. Different signals such as single spin flips yield similar results.

\section{Boundary Reflections}
\label{sec:reflections}
In this appendix we consider the case of signals impacting the boundaries of a non-moving window for several different models. We study the time
evolution beyond the time where a signal reaches the boundaries, both with Method I and Method II.
In \textit{all} cases we observe reflections from the boundary after some time.
The nature of these reflections generally depends on the boundary update method as well as the initial uniform state and the type of
the signal.

The models and signals that have been studied in particular are the TFI model with a JW excitation and a single spin flip in $x$-direction,
the XXZ model with a JW excitation and a single spin flip in $z$-direction, 
the $S=1$ Heisenberg model with a spin up excitation (this particular case is also studied in \cite{phien-infinite} with a method similar to Method II, but only for shorter times),
and the $S=1$ AKLT model \cite{AKLT} with a spin up excitation. We observe reflections from the boundary after some time in \textit{all} cases.

In the following we show results for the two cases of the TFI model with a JW excitation and the $S=1$ AKLT model  with a spin up excitation, where we have used Method II for the left boundary and Method
I for the right boundary to see their respective behaviour.
\subsection{TFI model with JW excitation}

  We again consider the TFI model at $h=0.45$ after a JW excitation in the infinite system ground state. 
We use a non-moving window with $N=50$ and maximum bond dimension $m_{\rm max}=120$, where the ground state MPS
representation has bond dimension $m_{0}=30$. The time evolution of the bipartite entanglement entropy $S_{\rm ent}(n,t)$
and the magnetization $S^{x}(n,t)$ can be seen in \Fig{fig:TFI_JW_absorb}. The signal reaches the boundaries
at $t\approx40$ and reflections start to emerge at $t\approx90$.

\begin{figure}[t]
 \centering
 \includegraphics[width=\linewidth,keepaspectratio=true]{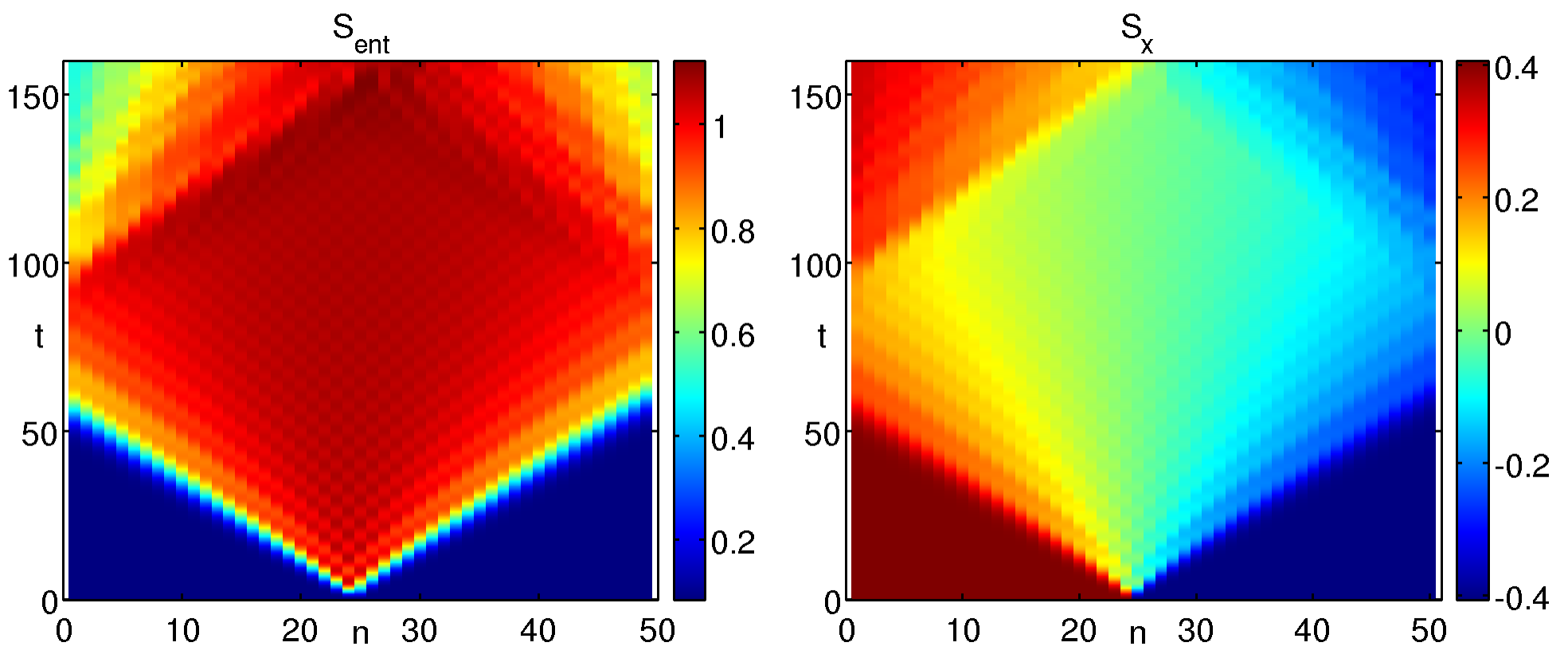}
 \caption{Time evolution of the bipartite entanglement entropy $S_{\rm ent}(n,t)$ (left) and the magnetization
$S^{x}(n,t)$ (right) in the TFI model at $h=0.45$ after a JW excitation within a non-moving CMW of size $N=50$. Time
evolution is continued after the signal has impacted the boundaries at $t\approx40$. Reflections 
emerge at $t\approx90$ at both boundaries, where we use Method I at the right and Method II at the left boundary.}
 \label{fig:TFI_JW_absorb}
\end{figure}

\begin{figure}[t]
 \centering
 \includegraphics[width=\linewidth,keepaspectratio=true]{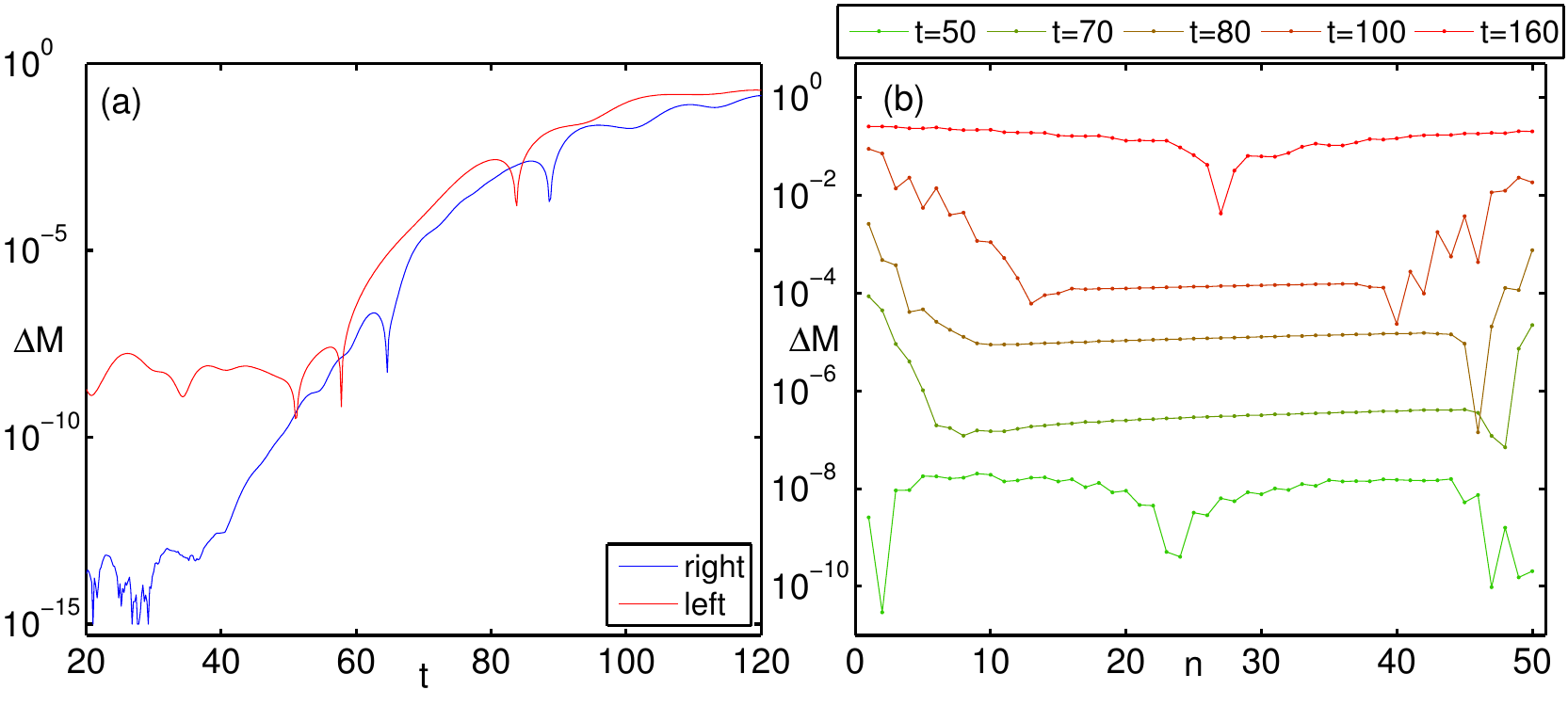}
 \caption{Difference $\Delta M(n,t)=\left|S^{x}(n,t) - S^{x}_{\rm ref}(n,t)\right|$ between
magnetization $S^{x}(n,t)$ of \Fig{fig:TFI_JW_absorb} and magnetization $S^{x}_{\rm ref}(n,t)$ of the
reference simulation of \ref{sec:precision}. (a) Difference at the immediate left and right boundaries of the $N=50$
non-moving window ($n=1$ and $n=50$ respectively) vs. time $t$. (b) Difference $\Delta M(n,t)$ vs. position $n$ inside
the non-moving window for various times $t$.}
 \label{fig:TFI_JW_absorb_compare}
\end{figure}

We compare the magnetization $S^{x}(n,t)$ of this simulation with the magnetization
$S^{x}_{\rm ref}(n,t)$ of the reference simulation of \ref{sec:precision} and show their absolute
difference $\Delta M(n,t)=\left|S^{x}(n,t) - S^{x}_{\rm ref}(n,t)\right|$ in
\Fig{fig:TFI_JW_absorb_compare}, where subplot (a) shows $\Delta M(n,t)$ at the left and right boundaries of the $N=50$
non-moving window ($n=1$ and $n=50$ respectively) vs. time $t$ and subplot (b) shows $\Delta M(n,t)$ vs. position $n$
inside the non-moving window at various
times
$t$.

In \Fig{fig:TFI_JW_absorb_compare}(a) it can be seen that initially the deviations at the right side (Method I) are
much lower than at the left side (Method II) until $t \approx 50$.
The deviation at both boundaries then increases exponentially
further until $t\approx100$, where it becomes of the order $\Or(1)$. We notice that the deviations for the right
boundary are always a bit lower than for the left boundary. We conclude that for the investigated case Method I performs slightly better than
Method II in absorbing a signal for a limited time.

\subsection{AKLT model with spin up excitation}

\begin{figure}[ht]
 \centering
 \includegraphics[width=\linewidth,keepaspectratio=true]{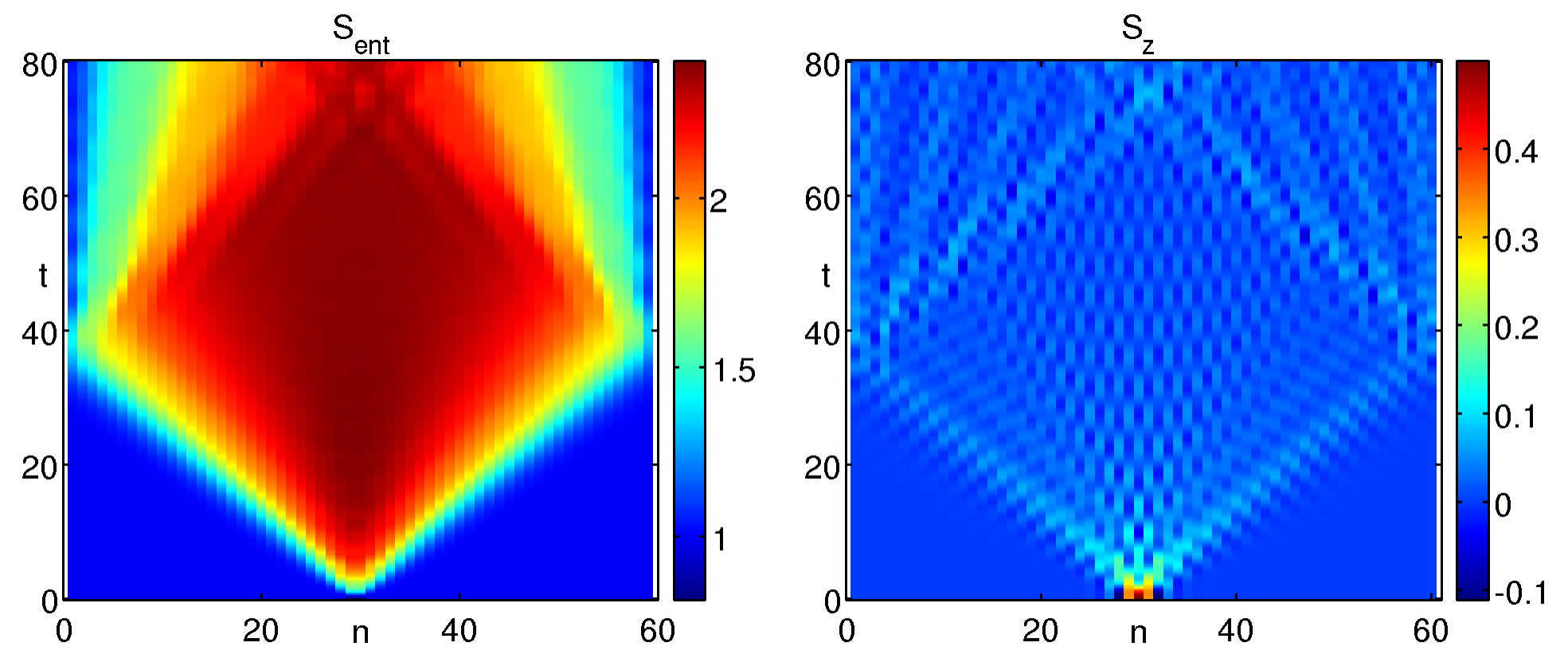}
 \caption{Time evolution of the bipartite entanglement entropy $S_{\rm ent}(n,t)$ (left) and the magnetization
$S^{z}(n,t)$ (right) in the AKLT model after an excitation induced by $\hat{S}^{+}_{n_{0}}$ on top of the infinite
system ground state within a non-moving CMW of size $N=60$. Time evolution is continued after the signal has impacted
the boundaries at $t\approx35$ to follow reflections which emerge almost immediately.}
 \label{fig:AKLT_absorb}
\end{figure}
  
  We also consider the $S=1$ bilinear, biquadratic chain at the \textit{AKLT} point \cite{AKLT} defined by the
Hamiltonian 
\begin{equation}
 \hat{H}=\sum_{j}\hat{\mathbf{S}}_{j}\cdot\hat{\mathbf{S}}_{j+1} +
\frac{1}{3}(\hat{\mathbf{S}}_{j}\cdot\hat{\mathbf{S}}_{j+1})^2.
\end{equation} 
The ground state is a valence bond state and has an exact MPS representation with bond dimension $m_{0}=2$ (see e.g.
\cite{MPS2}). We induce a signal on top of the infinite system ground state by applying the spin ladder operator
$S^{+}_{n_{0}}$. We use a non-moving window with $N=60$ sites and
maximum bond dimension $m_{\rm max}=200$. The time evolution of the bipartite entanglement entropy $S_{\rm ent}(n,t)$
and the magnetization $S^{z}(n,t)$ can be seen in \Fig{fig:AKLT_absorb}.

Here the signal impacting at $t\approx35$ is reflected almost immediately. This stems from the fact
that the MPS matrices at the boundary sites have to absorb all the information about excited states contained within the
propagating signal. Here these matrices however have bond dimension $m_{0}=2$ which is much too small for the matrices to
absorb this information for a long time span.


\end{document}